\documentclass[fleqn,usenatbib]{mnras}

\usepackage{newtxtext,newtxmath}

\usepackage[T1]{fontenc}
\usepackage{ae,aecompl}

\usepackage{graphicx}	
\usepackage{amsmath}	
\usepackage{hyperref}
\usepackage{natbib}
\usepackage[caption=false]{subfig}
\usepackage{ragged2e}
\usepackage{placeins}
\usepackage{bigints} 
\usepackage{physics}
\usepackage{xcolor}
\usepackage{cleveref}
\usepackage[export]{adjustbox}
\usepackage{longtable}
\usepackage{xtab}

\usepackage{scalerel,tikz}
\usetikzlibrary{svg.path}
\definecolor{orcidlogocol}{HTML}{A6CE39}
\tikzset{orcidlogo/.pic={
 \fill[orcidlogocol] svg{M256,128c0,70.7-57.3,128-128,128C57.3,256,0,198.7,0,128C0,57.3,57.3,0,128,0C198.7,0,256,57.3,256,128z};
 \fill[white] svg{M86.3,186.2H70.9V79.1h15.4v48.4V186.2z}
 svg{M108.9,79.1h41.6c39.6,0,57,28.3,57,53.6c0,27.5-21.5,53.6-56.8,53.6h-41.8V79.1z M124.3,172.4h24.5c34.9,0,42.9-26.5,42.9-39.7c0-21.5-13.7-39.7-43.7-39.7h-23.7V172.4z}
 svg{M88.7,56.8c0,5.5-4.5,10.1-10.1,10.1c-5.6,0-10.1-4.6-10.1-10.1c0-5.6,4.5-10.1,10.1-10.1C84.2,46.7,88.7,51.3,88.7,56.8z};
}}
\newcommand\orcidicon[1]{\href{https://orcid.org/#1}{\mbox{\scalerel*{
\begin{tikzpicture}[yscale=-1,transform shape]
\pic{orcidlogo};
\end{tikzpicture}
}{|}}}}

\def\sun{\odot}

\defcitealias{Mohapatra2020}{MFS20}

\title[Multiphase halos]{Characterising the turbulent multiphase halos with periodic box simulations}

\author[Mohapatra et al]{
Rajsekhar Mohapatra$^{\orcidicon{0000-0002-1600-7552}\,1}$\thanks{E-mail: rajsekhar.mohapatra@anu.edu.au (RM)},
Mrinal Jetti$^{\orcidicon{0000-0003-0287-3246}\,2}$\thanks{E-mail:mrinaljetti@gmail.com  (MJ)},
Prateek Sharma$^{\orcidicon{0000-0003-2635-4643}\,3}$\thanks{E-mail: prateek@iisc.ac.in (PS)} and
Christoph Federrath$^{\orcidicon{0000-0002-0706-2306}\,1,4}$\thanks{E-mail: christoph.federrath@anu.edu.au (CF)} 
\\
$^{1}$Research School of Astronomy and Astrophysics, Australian National University, Canberra, ACT~2611, Australia\\
$^{2}$ Department of Aerospace Engineering, Indian Institute Of Technology, Chennai, Tamil Nadu 600036, India\\
$^{3}$ Department of Physics, Indian Institute of Science, Bangalore, KA 560012, India\\
$^{4}$Australian Research Council Centre of Excellence in All Sky Astrophysics (ASTRO3D), Canberra, ACT~2611, Australia\\
}

\date{Accepted XXX. Received YYY; in original form ZZZ}

\pubyear{2021}

\begin{document}
\label{firstpage}
\pagerange{\pageref{firstpage}--\pageref{lastpage}}
\maketitle

\begin{abstract}
Turbulence in the intracluster medium (ICM) is driven by active galactic nuclei (AGNs) jets, by mergers, and  in the wakes of infalling galaxies. It not only governs gas motion but also plays a key role in the ICM thermodynamics. Turbulence can help seed thermal instability by generating density fluctuations, and mix the hot and cold phases together to produce intermediate temperature gas ($10^4$--$10^7$~$\mathrm{K}$) with short cooling times. We conduct high resolution ($384^3$--$768^3$ resolution elements) idealised simulations of the multiphase ICM and study the effects of turbulence strength, characterised by $f_{\mathrm{turb}}$ ($0.001$--$1.0$), the ratio of turbulent forcing power to the net radiative cooling rate. We analyse density and temperature distribution, amplitude and nature of gas perturbations, and probability of transitions across the temperature phases. We also study the effects of mass and volume weighted thermal heating and weak ICM magnetic fields. For low $f_{\mathrm{turb}}$, the gas is distribution is bimodal between the hot and cold phases. The mixing between different phases becomes more efficient with increasing $f_{\mathrm{turb}}$, producing larger amounts of the intermediate temperature gas. Strong turbulence ($f_{\mathrm{turb}}\geq0.5$) generates larger density fluctuations and faster cooling, The rms logarithmic pressure fluctuation scaling with Mach number $\sigma_{\ln{\bar{P}}}^2\approx\ln(1+b^2\gamma^2\mathcal{M}^4)$ is unaffected by thermal instability and is the same as
in hydro turbulence. In contrast, the density fluctuations characterised by $\sigma_s^2$ are much larger, especially for $\mathcal{M}\lesssim0.5$.
In magnetohydrodynamic runs, magnetic fields provide significant pressure support in the cold phase but do not have any strong effects on the diffuse gas distribution, and nature and amplitude of fluctuations.
\end{abstract}




\section{Introduction}\label{sec:introduction}
An interplay between different phases is seen in many different terrestrial and astrophysical systems across a vast range of scales. This includes interactions between water vapour and air in the earth's atmosphere \citep{Aronovitz1984PhRvA,Pal2016PhRvE}, the interface between hot and cold air in combustion \citep{Bray1991RSPSA}, solar corona mass ejections, loops, and coronal rain \citep{Foullon2011ApJ,Foullon2013ApJ,Antolin2020PPCF}, warm and hot phases of the interstellar medium (ISM) \citep{Begelman1990MNRAS,Slavin1993ApJ,Wolfire1995ApJISM,Vazquez-Semadeni2000ApJ,Audit2005A&A,Hennebelle2007A&A,Glover2010MNRAS}, 
neutral molecular/atomic, nebular and hot X-ray emitting regions in the Milky Way's wind \citep{Fox2015ApJ,Bordoloi2017ApJ,Teodoro2020Nature}, the circumgalactic medium (CGM) \citep{Wolfire1995ApJHVC,Tumlinson2011Sci,Werk2014ApJ,Tumlinson2017review,Marchal2021arXiv} and the intracluster medium (ICM) \citep{Tremblay2018ApJ,Gendron-Marsolais2018MNRAS,Olivares2019A&A,Boselli2019A&A,Rose2019MNRAS,Vantyghem2019ApJ,Vantyghem2021ApJ}.
The mixing between these different phases is often governed by turbulent gas motions. 
Understanding the nature and properties of these phases and 
the interactions between them is crucial for physical understanding.

In this study, we have focused on turbulence in the multiphase ICM of cool core clusters, where the ambient medium is made up of a hot X-ray emitting phase ($\sim\!10^7$--$10^8$~$\mathrm{K}$) with cooler 
atomic clouds/filaments, at $\sim\!10^4$~$\mathrm{K}$ often traced by H$\alpha$ emission, embedded in between. We also observe molecular gas at $\sim 10$~$\mathrm{K}$, traced by CO \citep{edge2001}, co-spatial with the atomic clouds \citep{hu1992,Conselice2001AJ,mcdonald2012,Werner2013ApJ,Werner2014MNRAS,Tremblay2018ApJ}. Turbulence in the intracluster medium is primarily driven by AGN jets and 
by mergers \citep{Balbus1990,Churazov2002MNRAS,Churazov2003,Omma2004MNRAS,Nelson2012ApJ}. In addition to mixing the hot and cold phases, turbulence also plays several other important roles in the ICM, such as heating and 
seeding density fluctuations.

Since the ambient ICM is hot, ionised and optically thin, it emits free-free bremsstrahlung and cools radiatively (see \citealt{Bohringer2010A&ARv} for a review), and the cooling time decreases with increasing density. Thus, cooler and denser gas cools faster, and without any heating, 
the cool core is expected to undergo runaway cooling flow 
until it eventually forms molecular gas. Turbulence can 
heat the ambient medium via turbulent dissipation, thus increasing its cooling time and averting a cooling flow. Turbulent mixing can transfer the AGN feedback power dissipated in the hot phase to the cooler gas. 
In these cases, turbulence prevents runaway cooling. 

Turbulence also seeds thermal instability, by generating density fluctuations in which the overdense regions cool faster. It can also mix the cold and hot phase gas together to make intermediate temperature gas ($10^5-10^6$~$\mathrm{K}$), which then cools rapidly into the cold phase. Turbulence affects the density and temperature distribution of the ICM gas. It is therefore important to accurately quantify the role of turbulence in the ICM.

Observational studies have helped constrain the temperature distribution of the hot and intermediate phase gas, 
in addition to the cooler phases mentioned above.  \cite{Bregman2006ApJ} have noted a lack of intermediate temperature gas at around $10^{5.5}$~$\mathrm{K}$, which is traced by O\texttt{vi}. \cite{Anderson2016MNRAS} have looked at the far-UV forbidden lines Fe\texttt{xxi} and Fe\texttt{xix} which roughly trace gas around $10^{6.5}$--$10^7$~$\mathrm{K}$. In general, the flux in the lowest X-ray and far UV temperatures is much lower than predicted by the cooling-flow models, as summarised in \cite{Peterson2006PhR}. The soft X-ray filaments are also known to be multiphase and are associated with $\mathrm{H}\alpha$ emission \citep{Sparks2004ApJ,Fabian2006MNRAS}.

Many recent observational studies have used different methods to estimate turbulent velocities in the ICM, summarised in the review by \cite{Simionescu2019SSRv}. The Hitomi space telescope \citep{hitomi2016} directly measured velocities of the hot-phase gas by 
resolving the line-broadening of Fe\texttt{xxv} and Fe\texttt{xxvi} lines and revealed low levels of turbulence (roughly $4$ percent of the thermal pressure) in the central regions of the Perseus cluster. XRISM\footnote{https://heasarc.gsfc.nasa.gov/docs/xrism}, the successor to Hitomi is expected to be launched in 2022, providing us a direct measurement of turbulent velocity in the volume-filling hot phase of the ICM for a sample of clusters. \citet{zhuravleva2014turbulent,zhuravleva2018} have measured X-ray surface brightness fluctuations of several nearby galaxy clusters and used it to indirectly infer velocities of the hot phase. \citet{khatri2016} have used  Planck data of the Coma cluster to calculate pressure fluctuations using the thermal Sunyaev-Zeldovich effect \citep{Zeldovich1969,Sunyaev1970Ap&SS} (tSZ), which can be further used to calculate turbulent pressure fluctuations and velocities.  More recently, \citet{Li2020ApJ} have used velocity measurements of gas in the cold ($10^4$~$\mathrm{K}$) and molecular ($10$~$\mathrm{K}$) phases in the ICM of nearby clusters to construct velocity structure functions. These structure functions are steeper than expected from Kolmogorov turbulence theory \citep{kolmogorov1941dissipation}.

Many recent numerical studies have also approached 
this problem, at many different scales. \cite{Ji2019MNRAS,Fielding2020ApJ,Tan2021MNRAS} have zoomed into the mixing layers between hot and cold phases. \cite{Armillotta2016,Wlady2016MNRAS,Gronke2018,Kanjilal2021MNRAS} and many others have looked at the survival of a cold cloud moving through a hot ambient medium. \cite{banerjee2014turbulence,Mohapatra2019,Grete2020ApJ} have looked at simulations of turbulence with thermal instability in idealised local box simulations. Studies such as \cite{Hillel2020ApJ,Wang2021MNRAS} have conducted isolated galaxy cluster simulations and looked at the turbulent velocity structure functions of hot and cold phases. Using similar setups, \cite{Wittor2020MNRAS} have looked at the relation between ensthropy, a proxy of turbulence and its sinks and sources. At even larger scales, 
\cite{Nelson2020MNRAS} have zoomed into individual halos in cosmological simulations and studied the multiphase environment within them.

While most of the small-scale simulations (mixing layer and wind-cloud) capture the interactions between the different phases in detail, they generally lack a more global perspective, such as the distribution of gas among different phases and statistical properties of the hot phase. Large cluster-scale and cosmological zoom-in simulations have enough samples of these multiphase interactions to get reliable statistics, but they often lack the resolution to resolve mixing layers and turbulence in much detail. Variable resolution in these simulations (due to adaptive mesh refinement, moving meshes or smoothed particle hydrodynamics) also makes it difficult to study turbulent statistics due to the spatio-temporal variation of numerical viscosity \citep{Fromang2007A&A,Mitran2009ASPC,Creasey2011MNRAS,Bauer2012MNRAS}. It is also difficult to do controlled parameter studies in these global simulations.

Local idealised box simulations are well-placed to tackle these issues - the boxes are large enough to perform these statistical studies, the setups are flexible and numerically cheap to do controlled parameter scans, while still having enough resolution to study turbulence through structure functions, power spectra and scaling relations between density, pressure and velocity fluctuations.
Among the recent studies of the multiphase turbulent ICM in these setups, \cite{banerjee2014turbulence,Mohapatra2019} lack the resolution to study turbulent structure functions. They also use a relatively high cutoff for the cold-phase gas ($10^6$~$\mathrm{K}$), which reduces the scale separation between different phases and may promote enhanced mixing between them. \cite{Grete2020ApJ} performed high-resolution magnetohydrodynamic (MHD) simulations, but they used an idealised cooling function (with proxies for free-free and linear cooling functions) to match the turbulent heating rate. These simulations, however, may have weaker cooling, since \cite{Mohapatra2019} using a similar setup showed that matching the turbulent heating rate to a realistic cooling rate would result in supersonic gas motions. The hot ICM is known to be subsonic. Hence to understand the interaction between the different phases and their kinematics, we need to conduct high-resolution simulations (with converged slopes of structure functions) with accurate implementation of 
cooling in the hot-phase gas, while maintaining the scale separation between the phases (by choosing lower cooling cutoffs, at $10^4$~$\mathrm{K}$).

Here we conduct a set of local simulations of homogeneous isotropic turbulence with radiative cooling in a box of size $40$~$\mathrm{kpc}$. We mainly scan the parameter space of the turbulent heating fraction parameter $f_{\mathrm{turb}}$ (defined in the next section). We also compare between our simulations with and without magnetic fields and two different types of idealised thermal heating models (namely heating $\propto\rho^1$ \& $\propto\rho^0$). In this study, 
 we introduce our model and setup. We mainly discuss the distributions of various thermodynamic properties (gas density, pressure and temperature) of the ICM. We study the statistical relations between these properties, such as the scaling relation between density and pressure fluctuations ($\sigma_s$ and $\sigma_{\ln{\bar{P}}}$) and the rms Mach number ($\mathcal{M}$). We analyse the phase diagrams of the ICM gas and the probability of transition between the different phases (using tracer particles), and how they vary with our parameters. In a companion study \citep{Mohapatra2021VSF}, we compare between the kinematics of the cold and hot phases of the gas in these simulations using velocity structure functions.

This paper is organised as follows. In \cref{sec:Methods}, we introduce our setup and numerical methods. In \cref{sec:results-discussion}, we present the key results of our simulations and discuss their implications. We discuss our caveats and future prospects in \cref{sec:caveats-future}. We finally summarise and conclude the study in \cref{sec:Conclusion}.

\section{Methods}\label{sec:Methods}
\subsection{Model equations}\label{subsec:ModEq}
We model the ICM as a fluid using the compressible MHD equations and the ideal gas equation of state. We solve the following equations:
\begin{subequations}
	\begin{align}
	\label{eq:continuity}
	&\frac{\partial\rho}{\partial t}+\nabla\cdot (\rho \mathbf{v})=0,\\
	\label{eq:momentum}
	&\frac{\partial(\rho\mathbf{v})}{\partial t}+\nabla\cdot (\rho \mathbf{v}\otimes \mathbf{v}+P^*I -\mathbf{B}\otimes \mathbf{B})=\rho\mathbf{F},\\
	\label{eq:energy}
	&\frac{\partial E}{\partial t}+\nabla\cdot ((E+P^*)\mathbf{v}-(\mathbf{B}\cdot\mathbf{v})\mathbf{B})=\rho\mathbf{F}\cdot\mathbf{v}+Q-\mathcal{L},\\
	\label{eq:induction}
	&\frac{\partial\mathbf{B}}{\partial t}-\nabla\times(\mathbf{v}\times\mathbf{B})=0,\\
	\label{eq:pressure}
	&P^*=P+\frac{\mathbf{B}\cdot\mathbf{B}}{2},\\
	\label{eq:tot_energy}
	&E=\frac{\rho\mathbf{v}\cdot\mathbf{v}}{2} + \frac{P}{\gamma-1}+\frac{\mathbf{B}\cdot\mathbf{B}}{2},
	\end{align}
\end{subequations}
where $\rho$ is the gas mass density, $\mathbf{v}$ is the velocity, $\mathbf{B}$ is the magnetic field, $P=\rho k_B T/(\mu m_p)$ is the thermal pressure, $\mathbf{F}$ is the turbulent force per unit mass that we apply, $E$ 
is the total energy density, $\mu$ is the mean molecular mass, $m_p$ is the proton mass, $k_B$ is the Boltzmann constant, $T$ is the temperature, $Q(t)$ and $\mathcal{L}(\rho,T)$ are the thermal heating and cooling 
rate densities respectively, and $\gamma=5/3$ is the adiabatic index. The cooling rate density $\mathcal{L}$ is given by
\begin{equation}\label{eq:cooling_function}
\mathcal{L} = n_en_i\Lambda(T),
\end{equation}
where $\Lambda(T)$ is the temperature-dependent cooling function of \cite{Sutherland1993} corresponding to $Z_{\odot}/3$ solar metallicity, and $n_e$ and $n_i$ 
are electron and ion number densities, respectively. Viscosity and thermal conduction are not included explicitly.

\subsection{Numerical methods}\label{subsec:numerical_methods}
We evolve equations \ref{eq:continuity} to \ref{eq:tot_energy} using the HLL5R Riemann solver \citep{Bouchut2007,Bouchut2010,Waagan2011} in a modified version of the FLASH code \citep{Fryxell2000,Dubey2008}, version 4. We use the MUSCL-Hancock scheme \citep{van1984SIAM,Waagan2009JCoPh} for time integration and a second-order reconstruction method that uses primitive variables and ensures positivity of density and internal energy. For magnetic fields, we use divergence cleaning in the form of the parabolic cleaning method of \cite{Marder1987JCoPh}. For most of our runs, we solve the hydrodynamic (HD) equations (by setting $\mathbf{B}=0$), but we also have two runs with MHD for comparison. We use a uniformly spaced 3D Cartesian grid with $L_x=L_y=L_z=L=40$~$\mathrm{kpc}$, where $L$ is the box size, with a default resolution of $384^3$. This gives us an effective spatial resolution of roughly $100$~$\mathrm{pc}$. Our boundary conditions are periodic. We have tested the code for convergence by doubling the resolution to $768^3$.

\subsubsection{Turbulent forcing}\label{subsubsec:Turb_forcing}
We follow a spectral forcing method using the stochastic Ornstein-Uhlenbeck (OU) process to model the turbulent acceleration field $\mathbf{F}$ with a finite autocorrelation time-scale $t_{\mathrm{turb}}$ \citep{eswaran1988examination,schmidt2006numerical,federrath2010}, which is fixed to $260~\mathrm{Myr}$ across all our simulations. We only excite the large-scale modes with $1\leq\abs{\mathbf{k}}L/2\pi\leq3$. The power is a parabolic function of $k$ and peaks at $k=2$ (we have dropped the wavenumber unit $2\pi/L$ for simplicity). For scales smaller than these injection scales ($k>3$), turbulence develops self consistently. We make sure that the driving field is solenoidal by removing the divergent component (component along $\mathbf{k}$). For more details of the forcing method, refer to section 2.1 of \cite{federrath2010}. We use the same acceleration field (with the same random seed) for all our simulations, except that we scale the amplitude dynamically to impose global thermal balance, which we describe in the following section.

\subsubsection{Global energy balance}\label{subsubsec:glob_energy_balance}
We maintain global energy balance in all our simulations, which is motivated by the lack of cooling flows in observations. 
We achieve this by controlling the amplitudes of thermal heating rate $Q$ and the turbulent energy injection rate.  We introduce a parameter $f_{\mathrm{turb}}$, which is the ratio of the turbulent energy injection rate to the radiative cooling rate. We impose the following condition at every time step:
\begin{equation}
    \int \rho\mathbf{F}\cdot\mathbf{v}\mathrm{d}V=f_{\mathrm{turb}}\int\mathcal{L}\mathrm{d}V.\label{eq:f_turb_scaling}
\end{equation}
The remaining energy loss is compensated by adding heat throughout the box. 
We follow two different methods 
- distributing this heat uniformly per mass and uniformly per volume. Mathematically, these are given by
\begin{subequations}
\begin{align}
    &Q_{\mathrm{mw}}(\mathbf{x},t)=\rho(\mathbf{x},t)(1-f_{\mathrm{turb}})\frac{\int\mathcal{L}\mathrm{d}V}{\int\rho\mathrm{d}V}\label{eq:Q_mw},\\
    &Q_{\mathrm{vw}}(\mathbf{x},t)=(1-f_{\mathrm{turb}})\frac{\int\mathcal{L}\mathrm{d}V}{\int\mathrm{d}V}.\label{eq:Q_vw}
\end{align}
\end{subequations}
Although both of these thermal heating methods are idealised, $Q_{\mathrm{mw}}$ is motivated by the several gas-density dependent heating processes, such as heating by photons and cosmic rays. Similarly, $Q_{\mathrm{vw}}$ represents the processes that deposit thermal energy in the hot volume-filling phase, such as AGN jets feeding buoyant bubbles, heating by mixing \citep{banerjee2014turbulence}.
We choose $Q=Q_{\mathrm{mw}}$ for most of our runs, but we also compare between these two implementations in \cref{subsubsec:diff_Q_model}.

\subsubsection{Cooling cutoff}\label{subsubsec:cool_cutoff}
Since we do not consider gravity in this setup, thermally unstable regions of gas can separate out from the hot phase and collapse to 
small scales \citep{field1965thermal,koyama2004field,sharma2010thermal}. Multiphase gas is prevented with gravity if the background $t_{\mathrm{cool}}/t_{\mathrm{ff}}\gtrsim10$ \citep{mccourt2012,choudhury2016}, where $t_{\mathrm{cool}}$ and $t_{\mathrm{ff}}$ are the cooling time and the free-fall time of the gas, respectively. In order to prevent the gas from cooling to very low temperatures, 
we set the cooling function to zero below a temperature $T_{\mathrm{cutoff}}$. Assuming that the cooling gas goes not fragment and conserves its mass 
and reaches pressure equilibrium with the hot phase, the scale of the smallest clouds scales as $T_{\mathrm{cutoff}}^{1/3}$. 
In addition to the temperature floor, we also impose a cooling pressure floor at 
roughly $P_0/600$, where $P_0$ is the initial pressure, for additional numerical stability 
and to prevent negative pressure. 
Thus, the cooling function takes the form
\begin{equation}
    \mathcal{L}=n_en_i\Lambda(T)\mathcal{H}(T-T_{\mathrm{cutoff}})\mathcal{H}(P-P_0/600), \label{eq:temp_pres_floor}
\end{equation}
where $\mathcal{H}$ is the Heaviside function. We set $T_{\mathrm{cutoff}}=10^4$~$\mathrm{K}$ for all of our simulations, which is also the lower temperature limit of the cooling table that we use \citep{Sutherland1993}. This choice is reasonable, since one would also need additional physics to model gas cooling below $10^4$~$\mathrm{K}$, such as accurate modelling of different species and chemical networks, and heating due to interstellar background radiation and cosmic rays. Moreover, even higher resolution would be needed to sufficiently resolve the densest gas.

\subsubsection{Relevant timescales}\label{subsubsec:time_scales}
Some important timescales of the system are the cooling time ($t_{\mathrm{cool}}$), thermal instability time scale $t_{\mathrm{TI}}$, sound crossing time ($t_{\mathrm{cs}}$) and turbulent mixing time ($t_{\mathrm{mix}}$). They are given by
\begin{subequations}
\begin{align}
    &t_{\mathrm{cool}}=\frac{E_{\mathrm{int}}}{\mathcal{L}}=\frac{P}{(\gamma-1)n_e n_i\Lambda(T)},\label{eq:t_cool}\\
    &t_{\mathrm{TI}}=\frac{\gamma t_{\mathrm{cool}}}{{2-\mathrm{d}\ln{\Lambda}/\mathrm{d}\ln{T}}-\alpha}, \label{eq:t_TI}\\
    &t_{\mathrm{cs}}=\frac{L}{\sqrt{\frac{\gamma k_B T}{\mu m_p}}}\text{, and} \label{eq:t_cs}\\
    &t_{\mathrm{mix}}=\frac{L_{\mathrm{inj}}}{v}\approx \frac{L}{2v}, \label{eq:t_mix}
\end{align}
\end{subequations}
where $Q\propto\rho^\alpha$, $\alpha=1$ for $Q_{\mathrm{mw}}$ and $\alpha=0$ for $Q_{\mathrm{vw}}$ and $v$ is the rms velocity.
For the gas in the hot phase ($T>10^7$~$\mathrm{K}$), $\Lambda(T)\propto T^{1/2}$, so $t_{\mathrm{TI}}\approx 3t_{\mathrm{cool}}$ for mass-weighted heating and $t_{\mathrm{TI}}\approx t_{\mathrm{cool}}$ for volume-weighted heating. See appendix of \cite{sharma2010thermal} for a derivation of $t_{\mathrm{TI}}$ using linear stability analysis.

\subsubsection{Sub-cycling for cooling}\label{subsubsec:cool_subcycle}
For gas in the intermediate phase ($2\times10^4$--$10^6$~$\mathrm{K}$), $t_{\mathrm{cool}}$ can be very short where $\Lambda(T)$ peaks ($\approx10^{5.5}$~$\mathrm{K}$). For an isobaric collapse, $n_e$, $n_i\propto1/T$, but since the hot-phase gas  is more volume-filling, $t_{\mathrm{cool}}$ is short only in 
small volume with high density and low temperature. Thus, it is numerically expensive to evolve the entire simulation domain at small time-steps set by the global minimum $t_{\mathrm{cool}}$ (given by $t_{\mathrm{cool,min}}$). Hence, we evolve the internal energy with operator splitting in the short $t_{\mathrm{cool}}$ grid cells 
using smaller time steps, such that the internal energy is not changing by more than $1/8$th per subcycle. The number of subcycles is dependent on the local $t_{\mathrm{cool}}$ at each cell, which can be different for different cells.  We evolve the rest of the cells at a slightly 
longer Euler time step ($\mathrm{dt}_{\mathrm{code}}$), such that it is 
always less than 5 times $t_{\mathrm{cool,min}}$  ($\mathrm{dt}_{\mathrm{code}}=\mathrm{min}(5t_{\mathrm{cool,min}},\mathrm{dt}_{\mathrm{CFL}})$, where $\mathrm{dt}_{\mathrm{CFL}}$ is the time step set by the Courant criterion). We have tested our code both with and without 
subcycling, and our results are not affected by it. 

\subsubsection{Tracer particles}\label{subsubsec:tracer_part}
We introduce 1000 Lagrangian tracer particles distributed uniformly throughout the volume at $t=0$. These tracers move around the simulation domain according to the velocity of the cell they are in. They track the local temperature, density and magnetic field, and the entire trajectory of these tracers is stored as a time-series for further analysis. These tracers do not have any back-reaction on the gas.

\subsection{Initial conditions}\label{subsec:init_conditions}
We initialise the gas with a temperature $T_0=4\times10^6$~$\mathrm{K}$, $n_e=0.086\,\mathrm{cm}^{-3}$, which corresponds to a sound speed of $300$~$\mathrm{km/s}$. The gas has $t_{\mathrm{cool}}\sim 19$~$\mathrm{Myr}$, $t_{\mathrm{TI}}\sim 57$~$\mathrm{Myr}$  and $t_{\mathrm{cs}}\approx64$~$\mathrm{Myr}$ across the whole box. Since initially the gas motion is subsonic, $t_{\mathrm{mix}}>t_{\mathrm{cs}}>t_{\mathrm{TI}}$. Thus, the gas is thermally unstable and forms multi-phase gas. Note that we start with gas temperatures somewhat lower than typical ICM temperatures of $1$--$2\times10^7$~$\mathrm{K}$. We choose these thermally unstable regions so that we can generate multiphase gas and study the interactions between the hot and cold phases. This setup can be directly compared with the inner regions of cool core clusters and not the whole ICM. After multi-phase gas condensation, the heat lost 
via cooling heats the ambient hot phase gas to typical ICM temperatures in the steady state (since we impose global energy balance). In reality, the hot phase temperature is set by the depth of the gravitational potential well but we do not model this in our simulations.

For our MHD runs, we set the initial magnetic field with equal mean and rms components, in the absence of strong observational constraints on the field geometry. The mean field is set in the $z$ direction. The rms component is set as a power law, taking the form $\mathrm{B}_k\propto k^{1.25}$, for $2\leq k \leq 20$ to model small-scale dynamo growth motivated by the Kazantsev spectrum \citep{Kazantsev1968JETP}. The initial plasma beta $\beta$ (ratio of thermal to magnetic pressure) is $100$. 


\subsection{List of Simulation models}\label{subsec:list_of_models}
We have conducted 10 simulations, mainly scanning the parameter space of $f_{\mathrm{turb}}$, which are listed in \cref{tab:sim_params} along with some relevant simulation parameters. The run labels indicate the value of $f_{\mathrm{turb}}$. We label the run \textbf{f0.10} as our \textbf{fiducial} run. Most of these runs use HD equations unless indicated in the label. Similarly, the default resolution is $384^3$ and the default model for thermal heating is $Q_{\mathrm{mw}}$ ($\propto \rho$; see \cref{subsubsec:glob_energy_balance}), unless indicated in the run label. For studying the effects of MHD, we introduce magnetic fields to the fiducial run. Similarly to study differences (if any) between the different models of thermal heating, we use volume-weighted thermal heating in f0.10vw. In order to check convergence of our results, we have higher resolution runs (labelled `HR', with $768^3$ resolution elements) for the fiducial, MHD and vw runs. We have run most of our simulations till $1.302$~$\mathrm{Gyr}$, and only two high-resolution simulations (f0.10HR and f0.10magHR) till $1.003$~$\mathrm{Gyr}$ since they are numerically expensive.

\begin{table*}
	\centering
	\caption{Simulation parameters for different runs.}
	\label{tab:sim_params}
	\resizebox{\textwidth}{!}{
		\begin{tabular}{lccccccc} 
			\hline
			Label & Resolution & $f_{\mathrm{turb}}$ & Thermal heating &$\mathcal{M}_{\mathrm{hot}}$ & $v$ $(\mathrm{km/s})$ & Magnetic fields  & $t_{\mathrm{end}}$ $(\mathrm{Gyr})$\\
			(1) & (2) & (3) & (4) & (5) & (6) & (7) & (8)\\
			\hline
			f0.001 & $384^3$ & $0.001$ & mass-weighted & $0.087\pm0.002$ & $110\pm5$ & N & $2.604$\\
			f0.04 & $384^3$ & $0.04$ & mass-weighted & $0.43\pm0.01$ & $320\pm30$ & N & $1.302$\\
			f0.10 (fiducial) & $384^3$ & $0.10$ & mass-weighted & $0.55\pm0.03$ & $350\pm40$ & N & $1.302$\\
			f0.10HR & $768^3$ & $0.10$ & mass-weighted & $0.91\pm0.05$ & $370\pm40$ & N & $1.003$\\
			f0.50 & $384^3$ & $0.50$ & mass-weighted & $1.55\pm0.08$ & $400\pm60$ & N & $1.302$\\
			f1.00 & $384^3$ & $1.00$ & N & $1.64\pm0.07$ & $410\pm80$ & N & $1.302$\\
		\hline
			f0.10mag & $384^3$ & $0.10$ & mass-weighted & $0.39\pm0.02$ & $330\pm40$ & Y & $1.302$\\
			f0.10magHR & $768^3$ & $0.10$ & mass-weighted & $0.59\pm0.05$ & $330\pm40$ & Y & $1.003$\\
		\hline
			f0.10vw & $384^3$ & $0.10$ & volume-weighted & $0.19\pm0.01$ & $340\pm20$ & N & $1.302$\\
			f0.10vwHR & $768^3$ & $0.10$ & volume-weighted & $0.23\pm0.01$ & $370\pm30$ & N & $1.003$\\
		\hline
			
	\end{tabular}}
	\justifying \\ \begin{footnotesize} Notes: Column~1 shows the simulation label. The number following `f' denotes f$_{\mathrm{turb}}$, also shown in column~3, which is the ratio of turbulent energy input rate to the radiative cooling rate, described in \cref{eq:f_turb_scaling}. In column~2 we list the resolution of the simulations. The default resolution of all the runs is $384^3$ cells, unless indicated by the label `HR' for $768^3$ cells. Column~4 lists the type of thermal heating (volume-weighted or mass-weighted) implemented in the simulations (see eq.~\ref{eq:Q_mw} and \ref{eq:Q_vw}) which is by default set to mass-weighted, unless indicated in the simulation labels as `vw'. In column~5, we denote the steady state rms Mach number ($\mathcal{M}_{\mathrm{hot}}$) of the hot-phase gas (see \cref{subsubsec:dens_pres_fluc} for its description). In column~6, we show the steady state rms velocity ($v$) in $\mathrm{km/s}$  and in column~7, we denote whether magnetic fields are switched on. Runs with magnetic fields are labelled by `mag'. Finally, in column~8, we show the end time of a simulation in $\mathrm{Gyr}$.\end{footnotesize} 
	
\end{table*}
\section{Results and discussion}\label{sec:results-discussion}
In this section, we present our results and discuss their possible interpretations. We first focus on the effects of different levels of turbulence (by varying $f_{\mathrm{turb}}$) and then describe the effects of higher resolution, MHD and volume-weighted thermal heating.

\subsection{2D projection maps}\label{subsec:2D_proj}
\begin{figure*}
		\centering
	\includegraphics[width=2.0\columnwidth]{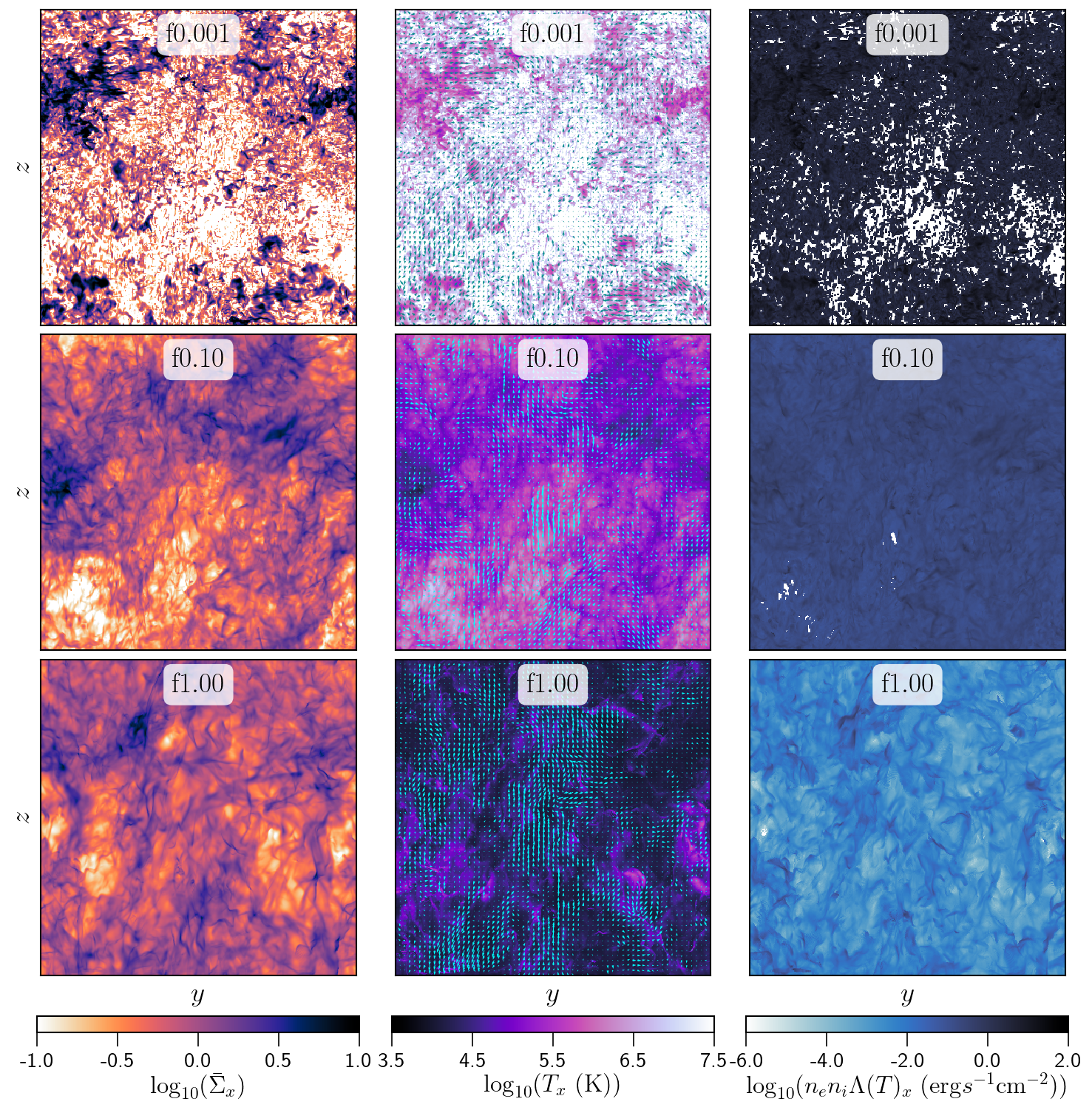}	
	\caption[density-temp-projection plots]{First column: snapshots of normalised density projected along the $x$ direction at $t=1.042$~$\mathrm{Gyr}$ for three representative simulations: f0.001, f0.10 and f1.00. Second column: mass-weighted projections of temperature at this snapshot along the $x$ direction for these runs. The cyan (dark-cyan for f0.001) arrows show the projected velocity field in the $yz$ plane. Third column: net emission from cold gas ($T<2\times10^4 \mathrm{K}$) along the $x$ direction. All colorbars are in log scale, and are shown at the bottom of each column. Note that the area covering fraction of cold gas is larger for higher turbulence, even though the mass fraction is larger for the f0.001 run as compared to the f1.00 run (see the top panel of \cref{fig:time-evolution}). These colormaps  are available in the cmasher package \citep{Ellert2020JOSS}.
	}
	\label{fig:dens-temp-proj-2d}
\end{figure*}
In \cref{fig:dens-temp-proj-2d} we show the line-of-sight (LOS) projections of density (column density $\Sigma_{\mathrm{LOS}}$, first column), mass-weighted temperature ($T_{\mathrm{LOS}}$, second column), and net emission from cold phase gas ($n_en_i\Lambda(T)_{\mathrm{LOS}}$, for $10^4$~$\mathrm{K}<T<2\times10^4$~$\mathrm{K}$, third column) for three representative simulations -- our fiducial run (f0.10, second row) and the two extreme runs with almost no-turbulent heating (f0.001, first row) and fully turbulent heating (f1.00, third row). We choose the $x$-direction as our LOS. In the second column, we also show the LOS-projection of the velocity field perpendicular to the LOS, as cyan arrows (dark-cyan for f0.001).  All the colorbars are in the log-scale. 

In the first column, for the f0.001 run, we observe that without turbulence, the gas is distributed bimodally into small dense clouds and large regions with almost no gas. They have gas densities around an order of magnitude larger and smaller than the mean density, respectively. In the fiducial run (second row), we observe a lot more gas at intermediate densities (orange-light purple colour). The size of the dense clouds is larger, and their densities are somewhat smaller. There are also less empty regions. In the third row, for the f1.00 run, the gas densities are even closer to the mean density, with the extreme density regions almost disappearing. Thus increasing turbulent heating fraction leads to increased mixing between the dense clouds and the ambient medium, which smoothens the density distribution. 

In the second column, for the no-turbulence run (f0.001), we notice that the lower temperature regions correspond to the dense clouds in the first column, which is expected.  The low density regions also correspond to the hot phase, which is at a fairly high temperature ($T\gtrsim10^{7.5}$~$\mathrm{K}$). For the fiducial run, we observe more gas at intermediate temperatures (between $2\times10^4$--$10^6$~$\mathrm{K}$). The hot phase is also at a slightly smaller temperature ($~10^7$~$\mathrm{K}$). For the f1.00 run, most of the 
projected temperature is close to or below the cooling cutoff at $10^4$~$\mathrm{K}$. We do not see a very clear hot phase, as there is not much gas with $T>10^6$~$\mathrm{K}$. 
For all of these runs, we do not observe any obvious correlation between the amplitude of projected velocity and the temperature/phase of the gas.

In the third column, we show the net emission measure from gas in the cold phase ($10^4$~$\mathrm{K}<T<2\times10^4$~$\mathrm{K}$), which is a proxy for H$\alpha$ emitting regions in the CGM/ICM. In the absence of turbulence, we observe emission coming from many small dense and cool clouds. With increasing turbulence, this emission gets smeared out. The magnitude of emission decreases but it covers many more sightlines along the LOS due to turbulent mixing. The emission strength for f1.00 run is also smaller because a lot of gas is cooled (below $10^4$~$\mathrm{K}$) due to strong turbulent rarefraction (see second column of third row), and such gas is assumed to be non-radiating. 

However, it is important to note that the size of the cold clouds could be much smaller and their area covering fraction much larger in reality (and in higher resolution simulations), since we do not resolve the cooling length ($\ell_{\mathrm{cool}}=c_st_{\mathrm{cool}}$) in our simulations. The scale $\ell_{\mathrm{cool}}$ reaches a minimum of $\approx0.1$~$\mathrm{pc}$ for the cold fast-cooling regions, 
much smaller than our resolution of $\approx100$~$\mathrm{pc}$. Later we have discussed the convergence of our results with respect to resolution 
(\cref{subsubsec:res_compare}). The appearance of smaller scale features with increasing resolution is expected as finer scales emerge with increasing Reynolds numbers. 

These results have implications for the observational probes of the CGM. The size and covering fraction of these cold clouds could be used as a crude estimate of the level of turbulence in these systems \citep{Tremblay2018ApJ,Olivares2019A&A,Vantyghem2021ApJ}.

\subsection{Evolution of density perturbations and energy }\label{subsec:time-evolution}
\begin{figure}
		\centering
	\includegraphics[width=\columnwidth]{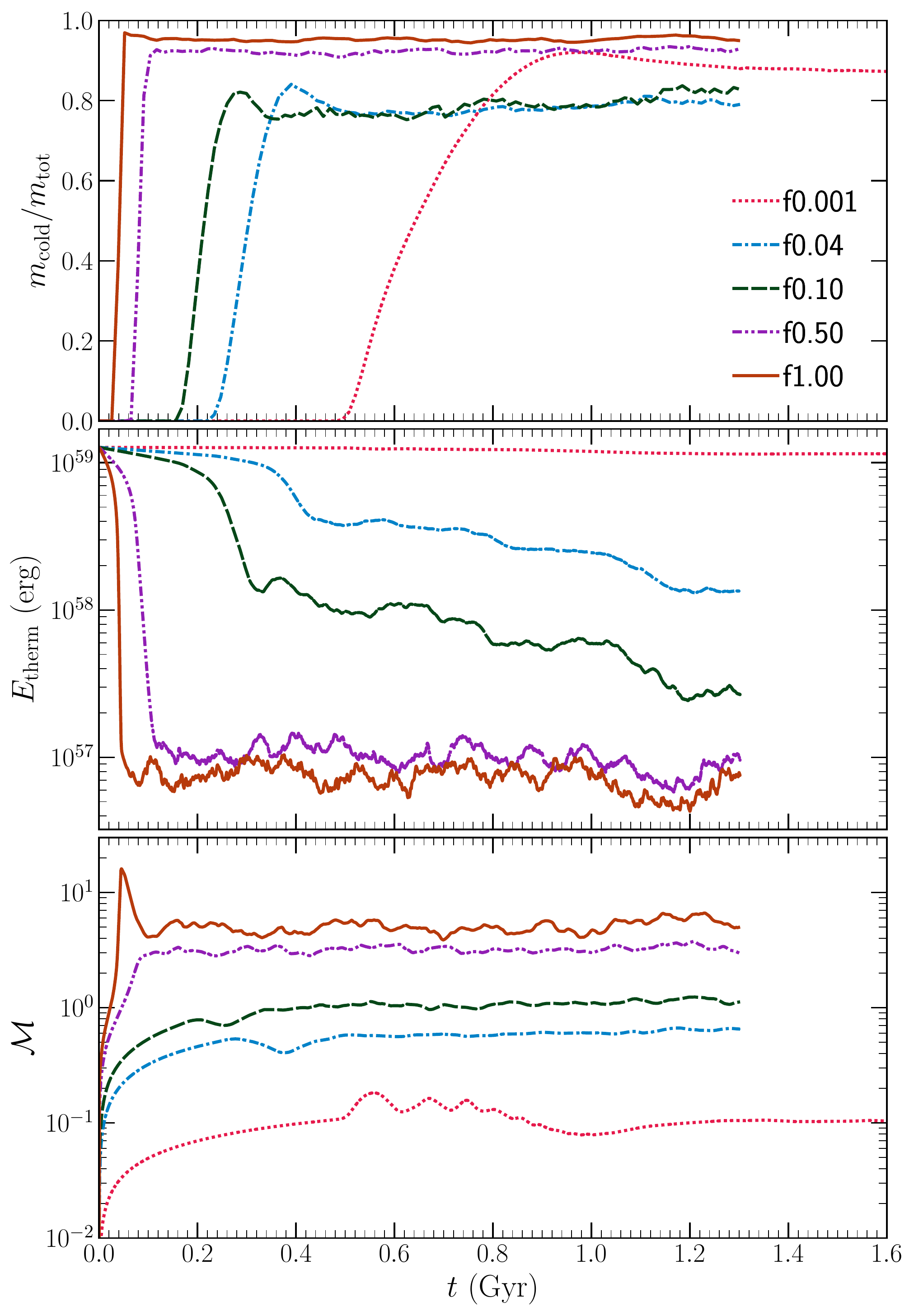}	
	\caption[time-evolution plots]{Time evolution of mass fraction of gas in the cold phase (first panel $m_\mathrm{cold}/m_{\mathrm{tot}}$),  thermal energy (second panel $E_{\mathrm{therm}}$), and volume-weighted rms Mach number (third panel $\mathcal{M}$). Although $E_{\mathrm{therm}}$ drops initially for the runs with stronger turbulence, the total energy (including  $E_{\mathrm{kin}}$) is approximately constant by construction. We have run the f0.001 run for longer time, since it reaches a steady state only around $t=1~\mathrm{Gyr}$.
	}
	\label{fig:time-evolution}
\end{figure}
In the upper panel of \cref{fig:time-evolution}, we show the net cold gas mass fraction (fraction of gas with $T<2\times10^4$~$\mathrm{K}$) as a function of time for our five 
runs with different $f_{\mathrm{turb}}$. In the middle panel, we show the time evolution of the thermal energy ($E_{\mathrm{therm}}$) and in the third panel we show the volume-weighted Mach number ($\mathcal{M}$). 

Our simulations start with the gas at $4\times10^6$~$\mathrm{K}$ (no cold gas). Turbulence generates density fluctuations in the gas. These regions are thermally unstable since $t_{\mathrm{TI}}<t_{\mathrm{mix}}$ ($t_{\mathrm{cool}}=19$~$\mathrm{Myr}$, $t_{\mathrm{TI}}\sim57$~$\mathrm{Myr}$, $t_{\mathrm{mix}}=92$~$\mathrm{Myr}$).
The over-dense regions cool all the way down to $10^4$~$\mathrm{K}$ (till $T_{\mathrm{cutoff}}$). We introduce a timescale $t_{\mathrm{multiphase}}$ which corresponds to the onset of cold gas in these simulations, defined as the time when $m_{\mathrm{cold}}/m_{\mathrm{tot}}>0.1$. This timescale $t_{\mathrm{multiphase}}$ is shorter for larger $f_{\mathrm{turb}}$ runs because they seed larger over-densities with shorter cooling times. 

In steady state, the value of $m_{\mathrm{cold}}/m_{\mathrm{tot}}$ (upper panel) first decreases with increasing $f_{\mathrm{turb}}$ till f0.10 and then increases. This can be attributed to the effects of turbulent mixing and turbulent rarefraction. For $f_{\mathrm{turb}}\lesssim0.1$, with increasing $f_{\mathrm{turb}}$, turbulent mixing between gas at the cooling cutoff and the hot phase gas pushes more and more of the cold gas to the intermediate phase (towards $\sqrt{T_{\mathrm{hot}}T_{\mathrm{cutoff}}}$ which decreases the mass fraction of cold gas with increasing $f_{\mathrm{turb}}$. But for $f_{\mathrm{turb}}\gtrsim0.1$, an increasing fraction of radiatively lost thermal energy is also 
supplied as kinetic energy because of the imposed global thermal balance, and this leads to supersonic turbulent velocities (see lower panel). This strong turbulence cools the hot phase gas (through supersonic rarefractions), which leads to an increase in $m_{\mathrm{cold}}/m_{\mathrm{tot}}$ with increasing $f_{\mathrm{turb}}$.

Since the simulations are in global energy balance, the lost $E_{\mathrm{therm}}$ is added back into the system in the form of turbulent kinetic energy ($E_{\mathrm{kin}}$, $f_{\mathrm{turb}}$ fraction) and thermal heat ($1-f_{\mathrm{turb}}$ fraction). The steady state ($t\gg t_{\mathrm{multiphase}}$) value of $E_{\mathrm{therm}}$ decreases with increasing $f_{\mathrm{turb}}$, and an increasing fraction of 
energy is retained as $E_{\mathrm{kin}}$. 
With increasing $f_{\mathrm{turb}}$, the effective conversion of thermal energy to $E_{\mathrm{kin}}$ also leads to increasing $\mathcal{M}$, as seen in the lower panel.

\subsection{Probability distribution functions}\label{subsec:PDF_dens}
In this section, we describe the probability distribution functions (PDFs) of various thermodynamic properties of our gas, and how they depend on the parameter $f_{\mathrm{turb}}$. This analysis can help us understand the observed CGM/ICM properties, such as the relative abundance/absence of intermediate temperature gas and how it depends on the strength of turbulence \citep{Werk2014ApJ,Tumlinson2017review} in the CGM and the ICM. They can also help us make predictions for future observations, such as quasar LOS studies of faint galaxies 
using the James Webb Space Telescope (JWST), emission from the hot phase gas using the Large Ultraviolet/Optical/Near Infrared Surveyor (LUVOIR\footnote{http://asd.gsfc.nasa.gov/luvoir/}) and the Advanced Telescope for High ENergy Astrophysics (ATHENA \footnote{http://sci.esa.int/cosmic-vision/54517-athena/}) .
\subsubsection{Mach number PDF}\label{subsubsec:mach_PDF}
\begin{figure}
		\centering
	\includegraphics[width=\columnwidth]{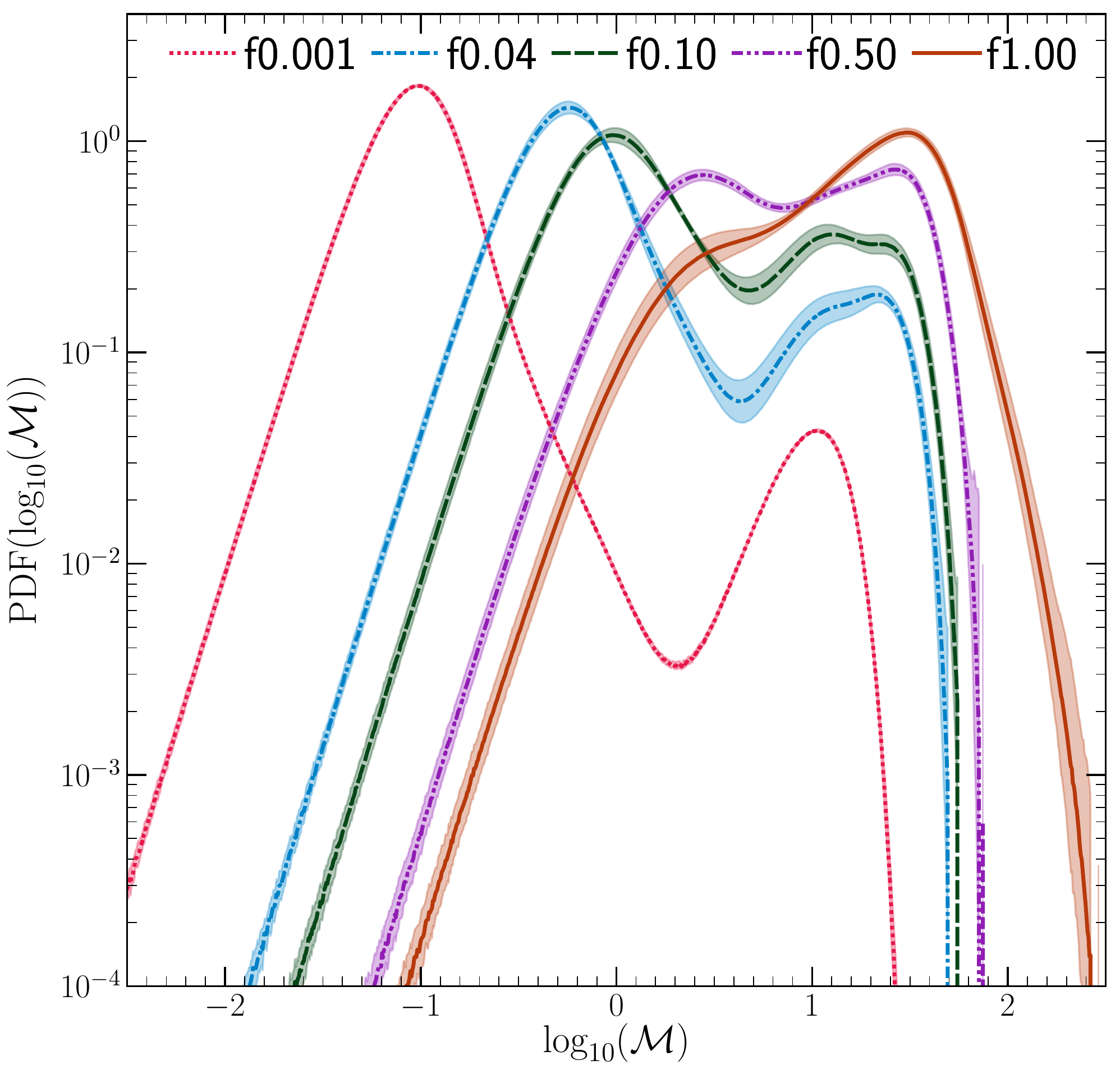}	
	\caption[Mach distribution plots]{The volume PDF of the logarithm of Mach number of the  gas for different $f_{\mathrm{turb}}$, averaged over 400 Myr in steady state. The shaded regions around the lines indicate the temporal variation (standard deviation of $\log_{10}(\mathrm{PDF})$). The higher Mach number peak becomes more prominent with increasing $f_{\mathrm{turb}}$ due to higher turbulent velocity and cooler gas. The hot phase peak (lower Mach number) becomes supersonic due to increased turbulent forcing.
	\label{fig:mach_PDF}
	}
\end{figure}
In \cref{fig:mach_PDF}, we show the volume PDF of Mach number. The PDFs have two distinct peaks, where the low Mach number peak corresponds to the hot phase (which has larger sound speed) and the high Mach number peak corresponds to the cold phase. As expected, with increasing $f_{\mathrm{turb}}$, both the peaks move towards larger Mach numbers due to stronger turbulent driving. The two peaks come closer to each other, and there is more gas at intermediate Mach numbers, due to increased turbulent mixing. The amplitude of the cold phase peak increases by an order of magnitude, since turbulent diffusion smears out the cold dense regions so that they occupy more volume. 

The rms velocity of the hot and cold phases is similar and the Mach number peaks essentially reflect the temperatures of the two dominant phases. The supersonic Mach number peak corresponding to the cold phase does not reflect the internal velocities within clouds but mostly their bulk motions.

The hot phase peak becomes supersonic for $f_{\mathrm{turb}}\gtrsim0.5$, which is inconsistent with the subsonic ICM \citep{hitomi2016}. In \cite{Mohapatra2019}, we used this argument to rule out turbulent dissipation to be the main source of heating the ICM, since it results in a larger turbulent to thermal pressure ratio compared to the observations by Hitomi (Table~\ref{tab:sim_params} shows that the mass-weighted thermal heating gives higher Mach numbers compared to the volume-weighted runs). However, the results from $f_{\mathrm{turb}}\gtrsim0.5$ could be applicable to smaller halos such as the CGM, which we discuss further in the following subsections.

\subsubsection{Temperature PDF}\label{subsubsec:temp_PDF}
\begin{figure}
		\centering
	\includegraphics[width=\columnwidth]{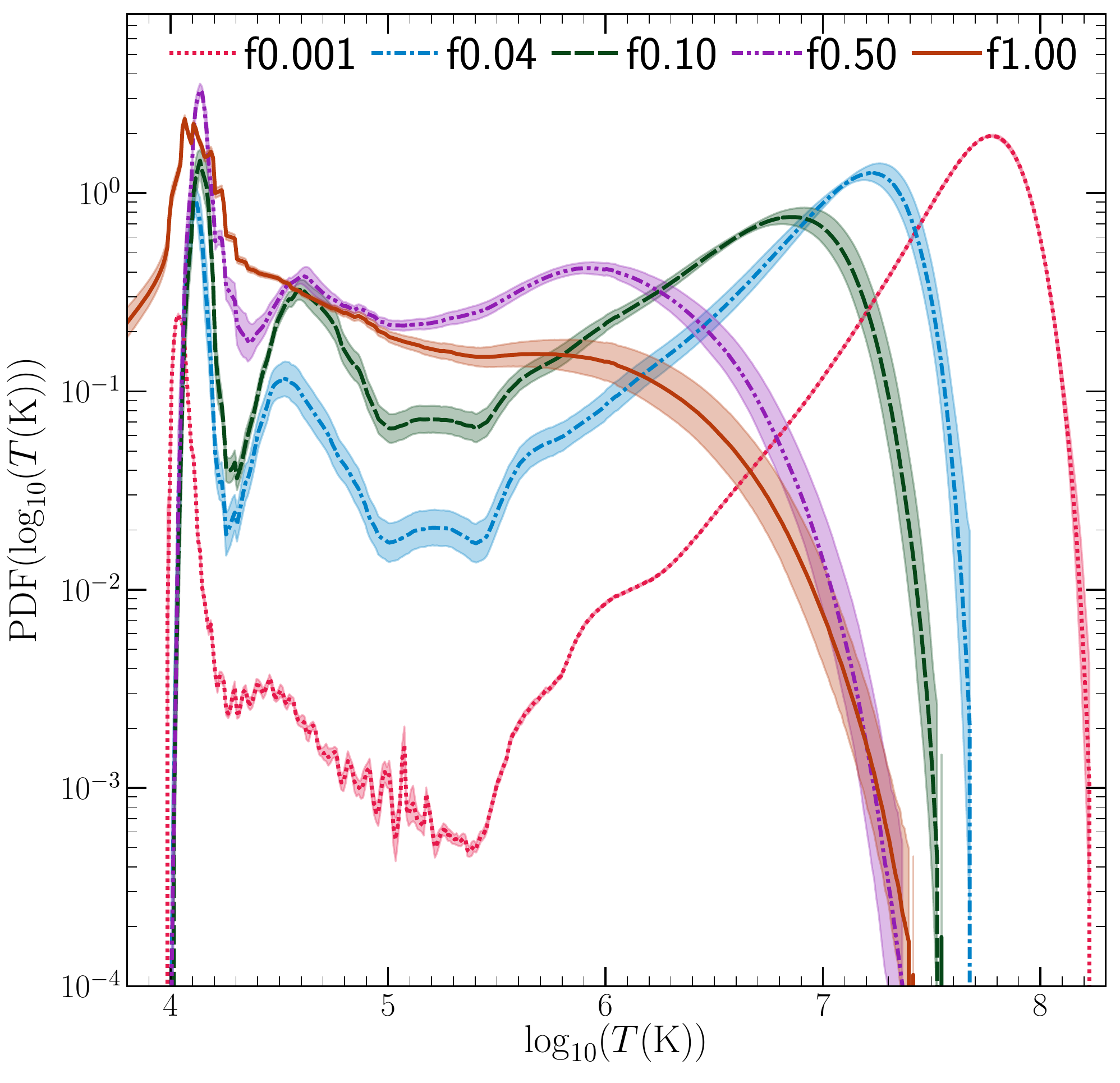}	
	\caption[Temperature distribution plots]{The volume PDF of the logarithm of temperature of the  gas for different fractions of turbulent heating. The amount of gas in the intermediate phase increases with increasing $f_{\mathrm{turb}}$. The hot phase is cooler for larger $f_{\mathrm{turb}}$ as turbulence mixes the hottest gas and cools it.
	\label{fig:temp_PDF}
	}
\end{figure}
We show the volume PDFs of temperature for different $f_{\mathrm{turb}}$ runs in \cref{fig:temp_PDF}. We observe two major peaks in most of the PDFs, one at $T_{\mathrm{cutoff}}$ corresponding to the cold phase and another between $10^6$--$10^8$~$\mathrm{K}$ at $T_{\mathrm{hot}}$, corresponding to the hot phase. Only for the f1.00 run, there is no peak at $T\gtrsim10^6$~$\mathrm{K}$. With increasing $f_{\mathrm{turb}}$, $T_{\mathrm{hot}}$ decreases, since a larger fraction of energy is deposited as turbulent energy rather than thermal. 
The volume of gas at intermediate temperatures also increases due to turbulent mixing between the hot and cold phases. For $f_{\mathrm{turb}}=0.04,0.10$ runs, the bumps in the PDF trace features of the cooling curve, with a dip near the cooling peak ($10^{5.5}$~$\mathrm{K}$, where $t_{\mathrm{cool}}$ is short) and the bumps correspond to build-up of gas just below this peak. For the f1.00 run, we also see a lot of gas below $T_{\mathrm{cutoff}}$. This gas cools due to rarefractions in supersonic turbulence.

The trends in the PDF with increasing $f_{\mathrm{turb}}$ are similar to the trends in temperature PDFs in \cite{Nelson2020MNRAS} with decreasing halo mass (see their figure 6). Smaller halos are 
expected to have 
larger deviations from hydrostatic equilibrium, and thus 
a larger turbulent Mach number \citep{Oppenheimer2018MNRAS}. The f0.50 and f1.00 temperature PDFs are almost flat in intermediate-hot phase temperatures due to highly efficient turbulent mixing. They appear similar to the temperature PDFs in the multiphase wind-cloud simulations in \cite{Kanjilal2021MNRAS} (see their figure 5). 

The abundance of gas at these intermediate temperatures is traced by ions such as Mg\texttt{II} ($\sim10^4~\mathrm{K}$), C\texttt{IV} and Si\texttt{IV} ($\sim10^{5}~\mathrm{K}$), N\texttt{V} and O\texttt{VI}($\sim10^{5.5}~\mathrm{K}$) (see fig.~6 of the review by \cite{Tumlinson2017review}), which can be detected through their absorption features in the spectra of background quasars and also by emission from nearby bright sources. Since the amount of gas at intermediate temperatures strongly depends on $f_{\mathrm{turb}}$, we can use the relative abundance of these ions to place constraints on the turbulent pressure fraction of the gas. Observations of galaxy clusters do not show strong, volume-filling emission in the far UV and soft X-ray bands, but rather find it to be confined to filamentary regions forming a boundary layer between the cold and hot phases \citep{Bregman2006ApJb,Werner2013ApJ,Anderson2018A&A}. These results favour a weak turbulent feedback (low $f_{\mathrm{turb}}$) scenario for the ICM.

\subsubsection{Pressure PDF}\label{subsubsec:pres_PDF}
\begin{figure}
		\centering
	\includegraphics[width=\columnwidth]{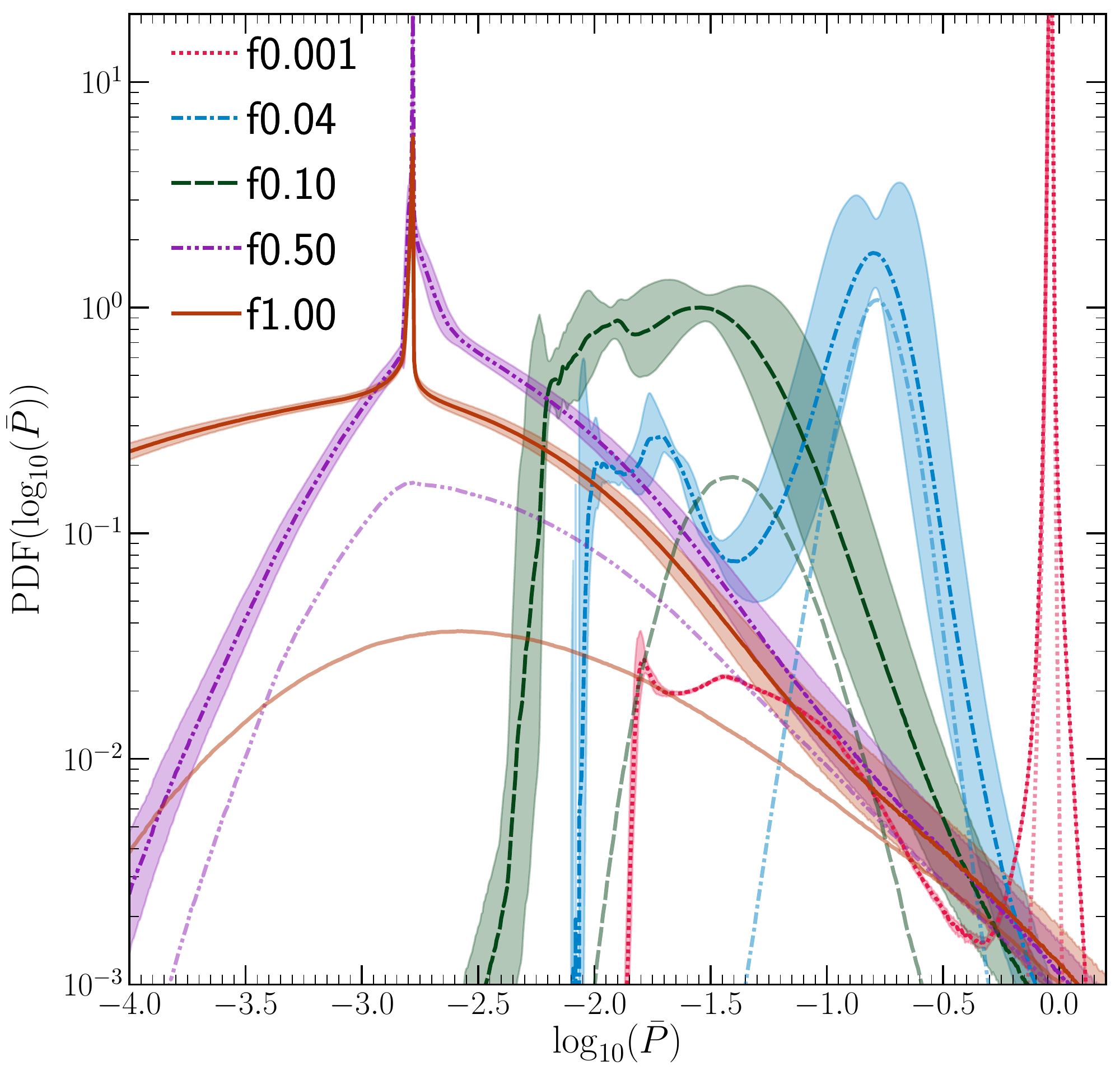}	
	\caption[Pressure distribution plots]{The volume-weighted PDF of the logarithm of pressure (normalised) for different fractions of turbulent heating, with shaded regions showing the $1$--$\sigma$ error interval. The lighter colored lines show the pressure PDFs for the hot phase gas ($T>10^7$~$\mathrm{K}$ for $f_{\mathrm{turb}}\leq0.10$ and $T>10^6$~$\mathrm{K}$ for $f_{\mathrm{turb}}>0.10$). A low pressure tail develops for the supersonic/transonic cases (f0.50 and f1.00 runs). The pressure PDF is bimodal for weaker driving, as gas at intermediate temperatures cools isochorically rather than isobarically. For the hot phase PDFs, note that the ambient pressure is smaller for larger $f_{\mathrm{turb}}$, which crudely mimics the CGM of different mass halos. \label{fig:pres_PDF}
	}
\end{figure}
In \cref{fig:pres_PDF}, we show the volume-weighted PDF of thermal pressure normalised by the initial mean value. The shapes of these PDFs vary the most with varying $f_{\mathrm{turb}}$. The low $f_{\mathrm{turb}}$ (f0.001, f0.04) runs show two peaks, where the low pressure peak corresponds to the cold phase. With increasing $f_{\mathrm{turb}}$, turbulence broadens these peaks (width $\propto\mathcal{M}^2$, see \citealt{Mohapatra2019,Mohapatra2021MNRAS}). The mean thermal pressure ($\langle P\rangle$) decreases with increasing $f_{\mathrm{turb}}$ because of predominant kinetic energy injection 
(also see second row of \cref{fig:time-evolution}).
The peak at $\log_{10}\bar{P}=-2.78$ in f0.50 and f1.00 runs corresponds to our cooling pressure floor (see eq.~\ref{eq:temp_pres_floor}), so gas with $P<P_{\mathrm{cutoff}}$ is generated by turbulent supersonic rarefactions.

The pressure distribution of the hot phase is relevant for tSZ observations, which is a measure of the LOS integral of the hot phase electron pressure. Here we define the hot phase as $T>10^7$~$\mathrm{K}$ for $f_{\mathrm{turb}}\leq0.10$ and $T>10^6$~$\mathrm{K}$ for $f_{\mathrm{turb}}>0.10$. This choice is reasonable since temperature PDFs of high $f_{\mathrm{turb}}$ runs are similar to smaller halos (intra-group medium and CGM) (see \cref{subsubsec:temp_PDF}), for which the hot phase is at $10^6$--$10^7$~$\mathrm{K}$.  

The hot phase pressure PDFs for these runs are shown in \cref{fig:pres_PDF} as lighter colored lines. For $f_{\mathrm{turb}}\leq0.10$, the pressure distribution is log-normal and its peak lines up with the high-pressure peak. The f0.50 and f1.00 runs show extended high-pressure tails and the PDF spans over $4$--$5$ orders of magnitude.

\subsubsection{Density PDF}\label{subsubsec:dens_PDF}

\begin{figure}
		\centering
	\includegraphics[width=\columnwidth]{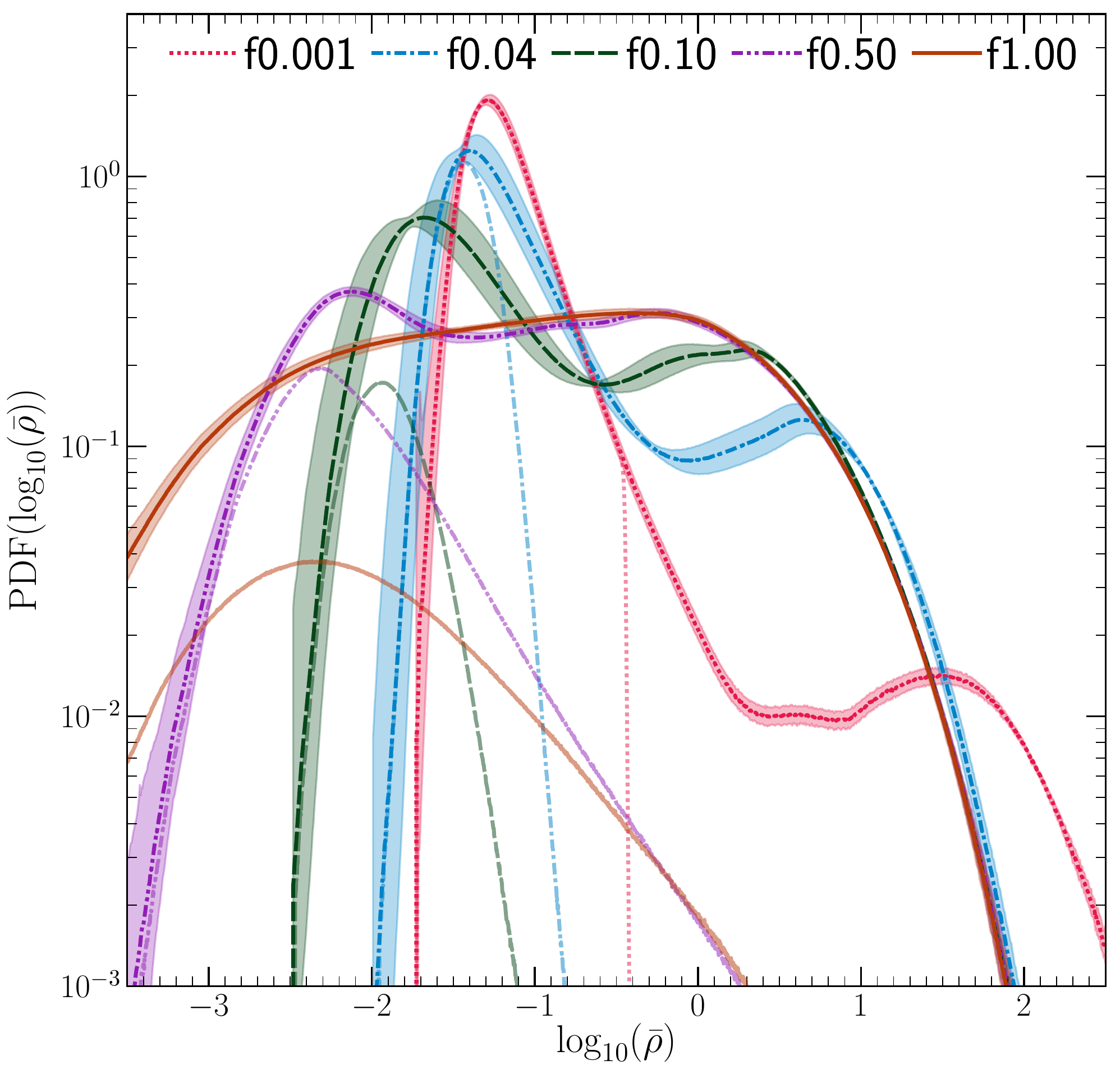}	
	\caption[Pressure distribution plots]{The volume PDF of the logarithm of density (normalised) of the  gas for different fractions of turbulent heating. The lighter coloured lines show the PDFs of gas in the hot phase ($T>10^7$~$\mathrm{K}$ for $f_{\mathrm{turb}}\leq0.10$ and $T>10^6$~$\mathrm{K}$ for $f_{\mathrm{turb}}>0.10$). The density bimodality decreases with increasing turbulence.
	\label{fig:dens_PDF}
	}
\end{figure}

We show the volume PDF of density (normalised by the initial value) for our different $f_{\mathrm{turb}}$ runs in \cref{fig:dens_PDF}. For $f_{\mathrm{turb}}<1.0$ we observe two peaks in the density distribution, where the denser phase corresponds to the cold phase and the rarer phase corresponds to the hot phase (see \cref{fig:dens-temp-proj-2d} for projection plots). The amplitude of the cold 
peak increases, whereas $\rho_{\mathrm{cold}}$ (the value of cold phase density peak) decreases with increasing $f_{\mathrm{turb}}$. This happens due to turbulent smearing of cold-phase gas---since the mass fraction of the cold phase gas is approximately similar across all runs (see first panel of \cref{fig:time-evolution}), with increasing turbulence, the cold regions are more spread out and have lower density. In other words, the dense regions are more fluffy for stronger turbulence.  

The density distribution of the hot phase 
is also observationally important, since 
it represents the X-ray emitting gas in the ICM. We show these PDFs as lighter colored lines in \cref{fig:dens_PDF}. The runs f0.04 and f0.10 have nearly log-normal distributions. The density distribution in all other runs (f0.001, f0.50 and f1.00) show a power-law tail at high densities. This happens because the high density gas cools faster, and its density increases as it cools, pushing it further to the right of the PDF. This shape of the PDF is also seen in the $\gamma=0.7$ runs in \cite{Federrath2015MNRAS} (see their figure 4).

\subsubsection{Density and pressure fluctuations as a function of $\mathcal{M}_{\mathrm{hot}}$}\label{subsubsec:dens_pres_fluc}
\begin{figure}
		\centering
	\includegraphics[width=\columnwidth]{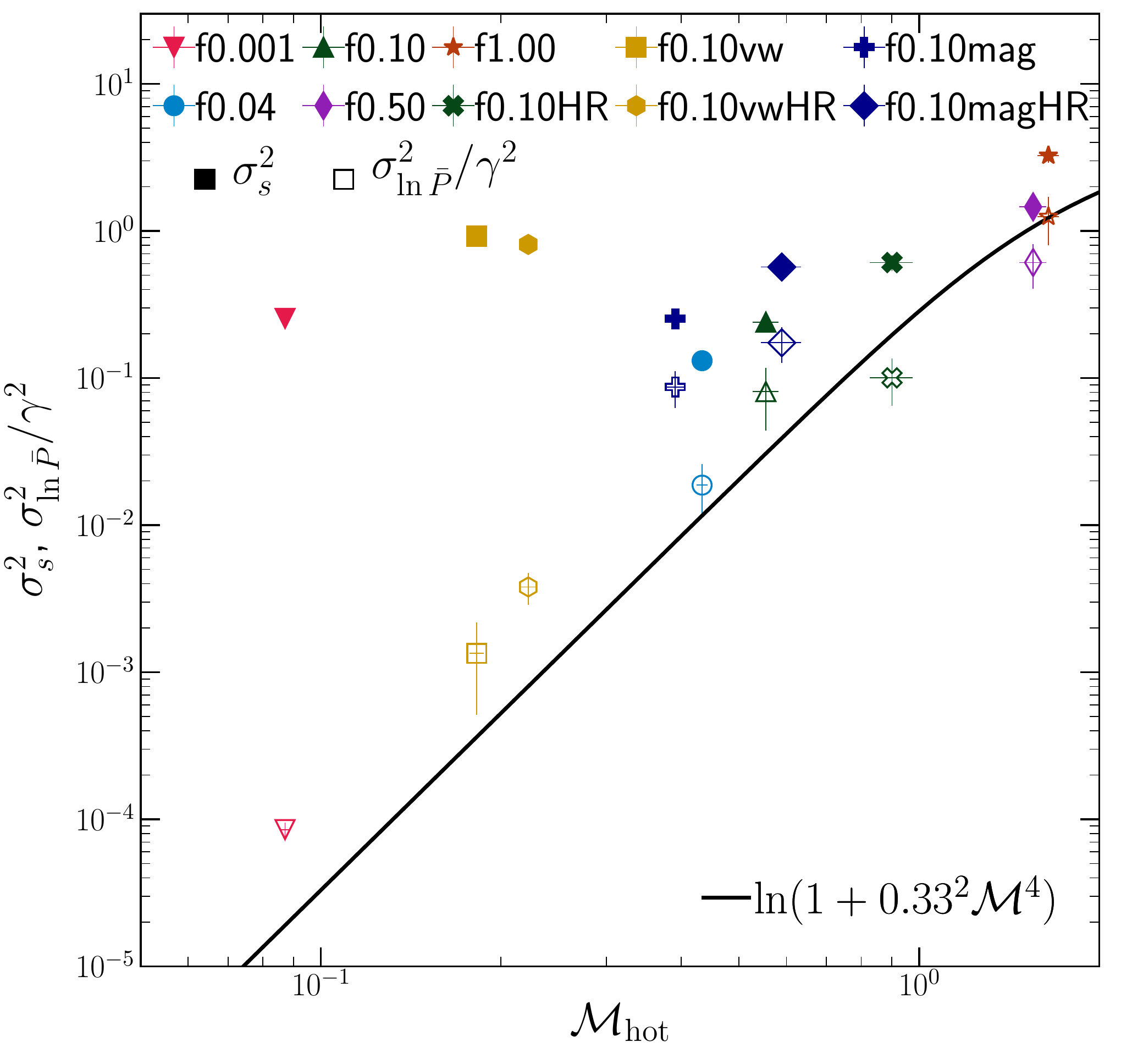}	
	\caption[hot gas density and pressure fluctuations]{The variance of the logarithm of normalised density ($s$, filled) and pressure ($\ln{\bar{P}}$, unfilled)  of the hot phase gas ($T>10^7$~$\mathrm{K}$ for $f_{\mathrm{turb}}\leq0.10$ and $T>10^6$~$\mathrm{K}$ for $f_{\mathrm{turb}}>0.10$) for all of our runs. The solid line shows the predicted scaling relation for subsonic turbulence with solenoidal driving and without radiative cooling (see eq.~13f in \citealt{Mohapatra2021MNRAS} with the Froude number $\mathrm{Fr}\rightarrow\infty$) and the driving parameter $b=0.33$ for solenoidal driving. Pressure fluctuations agree better with the scaling relation as compared to density fluctuations, which are much larger in the presence of radiative cooling and associated thermal instability.
	\label{fig:hot-sig-mach}
	}
\end{figure}

The density and pressure fluctuations in the hot phase gas in the ICM are used to obtain indirect estimates of turbulent gas velocities (see \cref{sec:introduction} and \citealt{zhuravleva2014turbulent,zhuravleva2014relation,khatri2016,Simionescu2019SSRv}). In \cref{fig:hot-sig-mach}, we show the scaling of the hot phase logarithmic density (filled data points) and logarithmic pressure fluctuations (unfilled) ($\sigma_s$ and $\sigma_{\ln{\bar{P}}})$) with the hot phase rms Mach number ($\mathcal{M}_{\mathrm{hot}}$) for our runs with different $f_{\mathrm{turb}}$. For convenience and ease of comparison, we have plotted $\sigma_{\ln{\bar{P}}}/\gamma$. We also show the $\sigma_s,\sigma_{\ln{\bar{P}}}/\gamma$--$\mathcal{M}$ scaling relation we proposed in \cite{Mohapatra2020,Mohapatra2021MNRAS} for density and pressure (here in the absence of gravitational stratification). We have used $b=0.33$ for solenoidal driving \cite[see eq.~23 in ][]{federrath2010}.

We denote the total density fluctuations as $\delta\rho_{\mathrm{tot}}$, and it has contributions from both turbulent density fluctuations ($\delta\rho_{\mathrm{turb}}=0.33\mathcal{M}^2$ for subsonic turbulence) and thermal instability ($\delta\rho_{\mathrm{TI}}$).
\begin{equation}
    \delta\rho_{\mathrm{tot}}^2=\delta\rho_{\mathrm{turb}}^2+\delta\rho_{\mathrm{TI}}^2 \label{eq:dens_fluc_total}
\end{equation}
Clearly, $\delta\rho_{\mathrm{TI}}^2$, (which is simply the difference between $\delta\rho_{\mathrm{tot}}^2$ and $\delta\rho_{\mathrm{turb}}^2$) has a much larger amplitude than $\delta\rho_{\mathrm{turb}}^2$ for $f_{\mathrm{turb}}\leq0.5$ (subsonic turbulence). For $f_{\mathrm{turb}} \gtrsim 0.5$ (also when turbulence becomes supersonic), $\delta\rho_{\mathrm{turb}}^2$ increases in magnitude, and dominates over $\delta\rho_{\mathrm{TI}}^2$.

The pressure fluctuations still follow the scaling relation with $\mathcal{M}_{\mathrm{hot}}$. In \cite{Mohapatra2020,Mohapatra2021MNRAS}, we showed that the pressure fluctuations are unaffected by stratification as well. This makes tSZ observations a more robust method for probing turbulent velocities than X-ray brightness fluctuations. With high angular resolution SZ observations (see \citealt{Mroczkowski2019} for a review), we can obtain reliable indirect estimates of turbulent velocities in the ICM. 

These results are similar to the trends reported in \cite{Mohapatra2019}, where the `with-cooling' runs had larger density fluctuations than the `without-cooling' runs (for the same $\mathcal{M}_{\mathrm{hot}}$), whereas the pressure fluctuations remained the same.

\subsection{Phase diagram of density and pressure fluctuations}\label{subsec:pres-dens-phase-diagram}
In this subsection, we present and discuss the joint 2D PDFs of pressure and density for three representative runs: f0.001, fiducial (f0.10) and f1.00. These phase diagrams 
can be used to characterise the mode of gas perturbations, which have been studied in thermal instability simulations \citep{Das2021MNRAS}, mixing layer turbulence \citep{Ji2019MNRAS,Fielding2020ApJ}, simulations with background stratification \citep{Mohapatra2020} and also in X-ray observations \citep{zhuravleva2018}. In this subsection, we discuss the role of different levels of turbulence and the onset of thermal instability on the nature of these perturbations.

\begin{figure*}
		\centering
	\includegraphics[width=2\columnwidth]{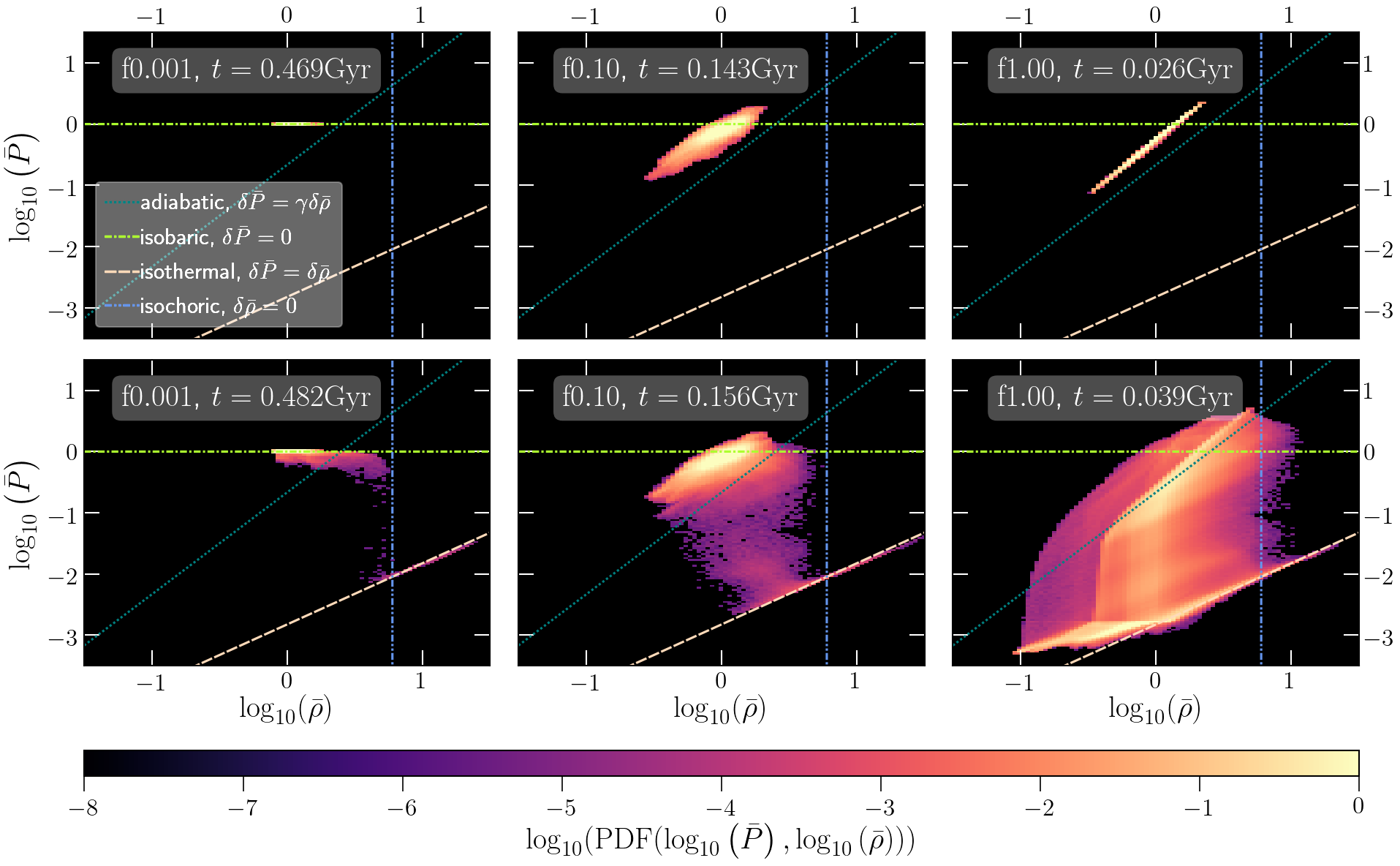}	
	
	\caption[multiphase pressure and density distribution plots]{The volume PDF of the logarithm of pressure versus logarithm of density for different fractions of turbulent heating just before (first row) and after (second row) multiphase gas formation for runs with different fractions of turbulent heating. The straight lines show the different fluctuation modes--adiabatic, isobaric, isothermal,  and isochoric. Before multiphase condensation, the lowest $f_{\mathrm{turb}}$ run is isobaric and the highest $f_{\mathrm{turb}}$ run is adiabatic. Notice the isochoric tracks at intermediate temperatures where $t_{\mathrm{cool}}\ll t_{\mathrm{cs}}$ across the cooling blob. Also notice the overdensity on the isothermal track (the `peach puff' coloured dashed line), which corresponds to $T_{\mathrm{cutoff}}$ at $10^4$~$\mathrm{K}$. The PDFs are broader with stronger turbulence.
	\label{fig:pres-dens-mp-2DPDF}
	}
\end{figure*}
In \cref{fig:pres-dens-mp-2DPDF}, we show these 2D-PDFs in two snapshots, 
just before (first row) and just after the formation of multiphase gas (second row). We have also shown different straight line fits which show different fluctuation modes--adiabatic, isobaric, isothermal,  and isochoric (see first column in upper panel).

Before the onset of thermal instability (first row), the fluctuation modes are isobaric for f0.001 run (first column) and with increasing $f_{\mathrm{turb}}$, the slope of the PDF increases and becomes adiabatic for f1.00 run. The amplitude of fluctuations is also much larger for larger $f_{\mathrm{turb}}$. 

For f0.001 run, only the highest density gas becomes multiphase and cools down to $T_{\mathrm{cutoff}}$ ($10^4$~$\mathrm{K}$). But for f0.10 and f1.00, large amplitude adiabatic modes lead to either gas with high density or lower density and lower temperature (closer to the peak of the cooling curve $\Lambda(T)$, see fig.~8 in \citealt{Sutherland1993} ). Both of these states are 
prone to condensation with $t_{\mathrm{TI}}<t_{\mathrm{mix}}$. Once $t_{\mathrm{cool}}\ll t_{\mathrm{cs}}$, the gas cools isochorically, where it overlaps with the isothermal fit ($T=T_{\mathrm{cutoff}}$). The overdensity at  $\log_{10}{\bar{P}}\approx-2.8$ in the third column of the second row corresponds to $P_{\mathrm{cutoff}}$ (see eq.~\ref{eq:temp_pres_floor}). 

\cite{Fielding2020ApJ} argue that the isochoric behavior is due to lack of sufficient resolution. While this is partly true, we expect large blobs to cool isochorically through $10^{5}$~$\mathrm{K}$, the peak of the cooling curve (e.g., see \citealt{Das2021MNRAS}). We present a resolution study of the 1D pressure PDFs in \cref{subsubsec:res_compare}.

Recent observations point towards the ICM being mainly dominated by isobaric modes (see fig.~6 in \citealt{zhuravleva2018}). This is in line with our results---stronger turbulence leads to larger, adiabatic fluctuations whereas weaker subsonic turbulence produces mainly isobaric modes in the hot phase. In \cite{Mohapatra2020}, 
we showed that gravitational stratification can also change the perturbation modes from adiabatic to isobaric.
 Our results agree with the cluster scale simulations of \cite{gaspari2013constraining}. \citeauthor{gaspari2013constraining} also showed that strong thermal conduction can introduce isothermal modes, but conduction may be suppressed in the ICM, e.g., as seen in a recent plasma experiment \citep{Meinecke2021arXiv}. 

\subsection{Tracer particles}\label{subsec:tracers}
In this subsection we present the time evolution and statistics of the Lagrangian tracer particles 
(see \cref{subsubsec:tracer_part}). These tracers are initially uniformly distributed throughout the volume and move with the local flow. They do not have any back-reaction on the fluid. 

\subsubsection{Evolution of particle properties}\label{subsubsec:tracer-evol}
\begin{figure}
		\centering
	\includegraphics[width=\columnwidth]{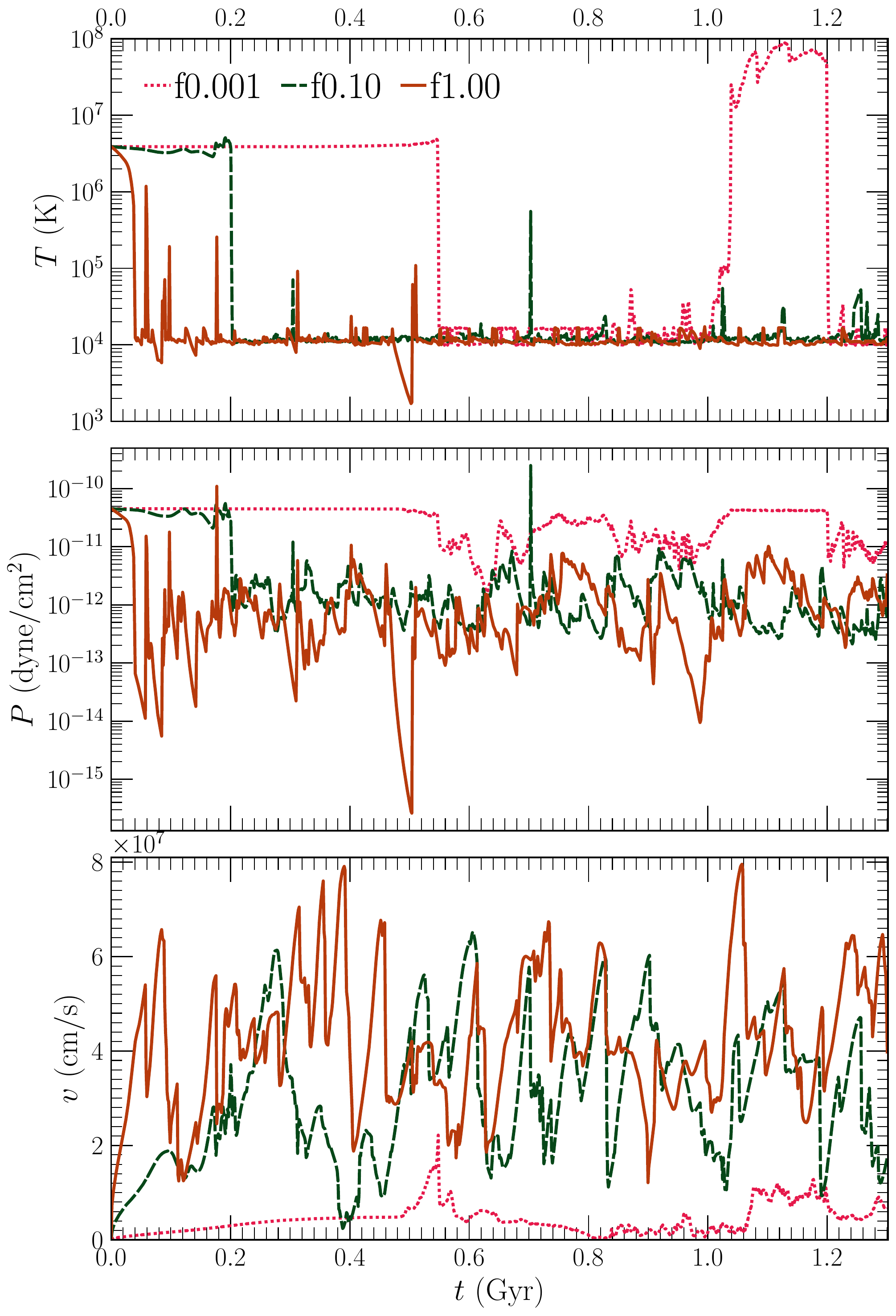}	
	\caption[tracer particle evolution plot]{The  time evolution of properties of a randomly chosen Lagrangian tracer particle: temperature (first row), pressure (second row) and magnitude of velocity (third row) for runs with different levels of turbulence. The number of excursions across phases increases with increasing $f_{\mathrm{turb}}$, as expected. Velocities also increase with larger $f_{\mathrm{turb}}$.
	\label{fig:tracers-evolution}
	}
\end{figure}
In \cref{fig:tracers-evolution}, we show the time evolution of $T$ (first row), $P$ (second row) and magnitude of velocity ($v$, third row) of a randomly chosen particle for three runs f0.001, fiducial (f0.10) and f1.00. All three particles are initially in the hot phase and then transition to the cold phase. Once the temperature of the particle drops, it  drops all the way to $T_{\mathrm{cutoff}}$, since the intermediate region is fast-cooling and short-lived. These phase changes are all associated with a drop/rise in $P$. We also observe that these transitions are not entirely isochoric---for example, at $t\approx0.2$~$\mathrm{Gyr}$, for the fiducial run, $T$ drops by 2.5 orders of magnitude, whereas $P$ only drops by 1.5 orders, implying that the density $\rho$ increases by an order of magnitude. The drop in $P$ during these transitions increases with increasing $f_{\mathrm{turb}}$. 

There are clear drops in the particle velocities for f0.001 run during hot-cold transition and a rise in the velocity during the cold-hot transition. But for the fiducial and f1.00 runs, the stochastic turbulent velocity changes are large and difficult to distinguish from the velocity changes, if any, associated with the phase transitions. 

\subsubsection{Trajectory of tracers}\label{subsubsec:tracer-trajectory}

\begin{figure*}
		\centering
	\includegraphics[width=2.0\columnwidth]{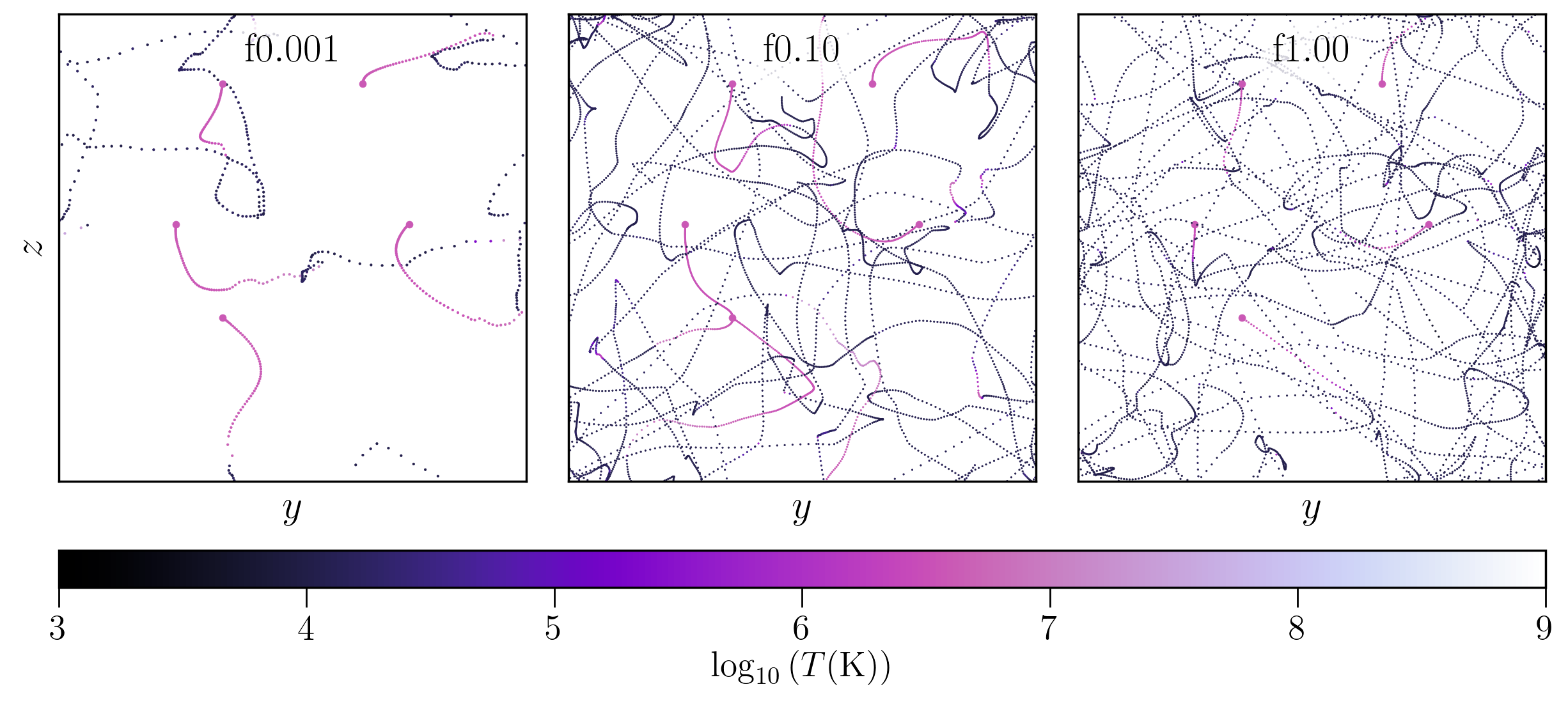}	
	\caption[tracer-temp-2D plots]{The projected trajectory in the $yz$~-plane of five randomly chosen tracer particles with same starting location over the entire duration of our simulations ($1.302$~$\mathrm{Gyr}$). The three panels are for our three representative simulations: f0.001, f0.010 and f1.00. The color represents the temperature of the tracer particle. The initial position of the particle is denoted by a large pink dot. The subsequent scatter points are $10.4$~$\mathrm{Myr}$ apart for f0.001 and $1.3$~$\mathrm{Myr}$ apart for runs f0.10 and f1.00. The runs with higher turbulence have much larger diffusion of the tracer particles, as seen by a larger coverage of the $yz$ plane.
	}
	\label{fig:tracer-trajectory}
\end{figure*}
Here we discuss the trajectories of tracer particles for our f0.001, fiducial and f1.00 runs (tracers for different runs have the same starting location). In \cref{fig:tracer-trajectory}, we show the projected trajectories along the $yz$ plane of five tracers, colored by their instantaneous temperature. The position and temperature of each tracer is shown every $10.4$~$\mathrm{Myr}$ for the f0.001 run and every $1.3$~$\mathrm{Myr}$ for the f0.10 and f1.00 runs. 

The area covering fraction of projected trajectories increases with increasing $f_{\mathrm{turb}}$ due to 
higher turbulent diffusion. The particles cover larger distances due to 
larger turbulent velocities. For f0.001 run, we observe that many phase transitions (change in color) are also associated with jerks in the trajectory, whereas there is no noticeable effect for larger $f_{\mathrm{turb}}$ runs, meaning hot and cold phases may be co-moving for stronger turbulence. This has important implications for observational studies such as \cite{Li2020ApJ}, where the authors measure velocities of the cold phase gas and use it to estimate velocities of the hot phase. However, note that in the presence of gravitational stratification, the cold phase may lose pressure support and move 
relative to the hot and roughly hydrostatic atmosphere \citep{Wang2021MNRAS}.

\subsubsection{Probability of phase transitions}\label{subsubsec:tracer-probability}

\begin{figure}
		\centering
	\includegraphics[width=1.0\columnwidth]{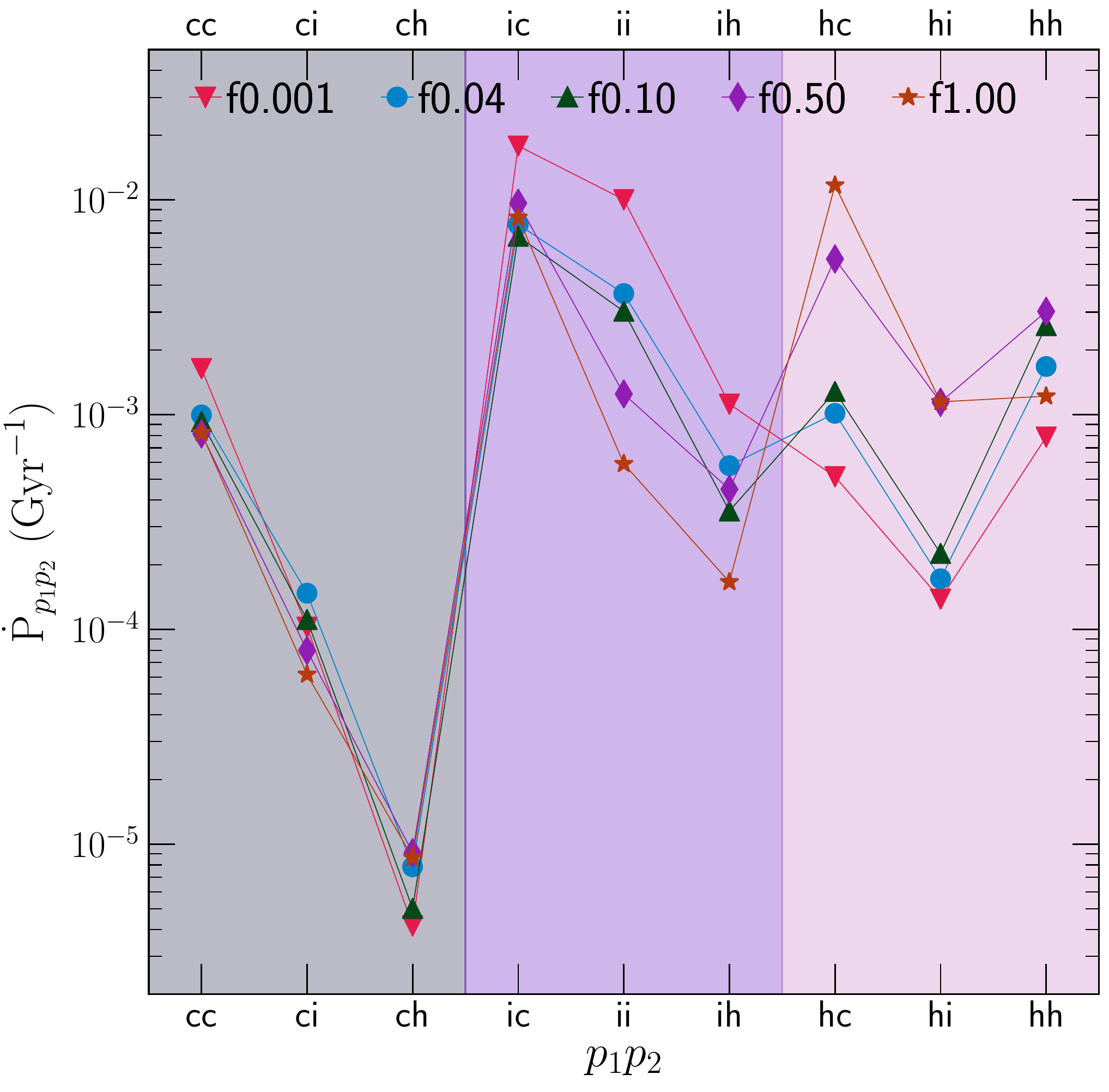}	
	\caption[tracer-scatter plots]{A scatter plot of transition probability per unit time of 
	particles from phase $p_1$ to $p_2$ (denoted by $\dot{P}_{p_1p_2}$).
	The background colour represents the phase $p_1$---grey, purple and pink for cold, intermediate and hot phases, respectively. Among the three phases, the particles are most likely to transition into the cold phase, and most likely to transition out of the intermediate phase. There are also clear trends with $f_{\mathrm{turb}}$, which we discuss in the main text.}
	\label{fig:tracer-transitions-scatter}
\end{figure}

The probability of phase transitions per time ($\dot{P}_{p_1p_2}$ for a transition from phase $p_1$ to $p_2$) is another interesting thermodynamic statistic---it gives us information about the relative stability of different phases and the amount of mixing between them. These are motivated by the Einsteins' coefficients for transition between different energy levels of a system. We divide the gas between three phases---cold (denoted by `c', $T<2\times10^4$~$\mathrm{K}$), intermediate (denoted by `i', $2\times10^4$~$\mathrm{K}<T<10^6$~$\mathrm{K}$) and hot (denoted by `h', $T>10^6$~$\mathrm{K}$)\footnote{Note that we define $T_{\mathrm{hot}}=10^6~\mathrm{K}$ across all our runs, unlike \cref{subsubsec:dens_pres_fluc}. This makes it more straightforward to compare between the different transitions probabilities as a function of $f_{\mathrm{turb}}$.}. We use a peak (and dip) detector algorithm to note changes in the temperature of the tracers, which we mark as phase changes. 
The quantity $\dot{P}_{p_1p_2}$ is defined as:
\begin{subequations}
\begin{equation}
    \dot{P}_{p_1p_2}=\frac{N_{p_1p_2}}{N_{p_1}t_{p_1}},\label{eq:prob_phase_transition}
\end{equation}
where $N_{p_1p_2}$is the total number of transitions from phase $p_1$ to $p_2$, $N_{p_1}$ is the total number of transitions starting from $p_1$ ($N_{p_1}=N_{p_1c}+N_{p_1i}+N_{p_1h}$), and $t_{p_1}$ is the total time spent by a particle in the phase $p_1$. The probability of transitioning out of the phase $p_1$ per unit time is denoted by $\dot{P}_{p_1}$ and is defined as:
\begin{align}
    &\dot{P}_{p_1}=\dot{P}_{p_1c}+\dot{P}_{p_1i}+\dot{P}_{p_1h}=1/t_{p_1}, \text{ so} \label{eq:prob_phase_out} \\
    &\int_{0}^{t_{p_1}}\dot{P}_{p_1}\mathrm{d}t=1.\label{eq:normalisation_transition}
\end{align}
\end{subequations}
So a larger value of $\dot{P}_{p_1}$ corresponds to less time spent in phase $p_1$, a more unstable phase. We average all these quantities over all (1000) tracers. 

We have shown $\dot{P}_{p_1p_2}$ for all possible phase transitions in \cref{fig:tracer-transitions-scatter}. The transition `ic' is most likely, as expected, since gas in the intermediate phase has the shortest cooling time and it cools rapidly to the cold phase. Transition from the cold phase to other phases `ci' and `ch' are least likely, since particles spend most of their time in the cold phase ($m_{\mathrm{cold}}/m_{\mathrm{tot}}\sim0.7$--$0.9$, see \cref{fig:time-evolution}). 
The rate $\dot{P}_i$ ($=\dot{P}_{ic}+\dot{P}_{ii}+\dot{P}_{ih}=1/t_{i}$) decreases with increasing $f_{\mathrm{turb}}$, since the amount of intermediate temperature gas increases, leading to a larger $t_i$ (see \cref{fig:temp_PDF}) due to increased turbulent mixing. The opposite trend is seen in the transition rates `hc' and `hi', which increase with increasing $f_{\mathrm{turb}}$, as the mean temperature of the hot phase decreases, leading to a shorter cooling time. However, $\dot{P}_{hi}\ll\dot{P}_{hc}$, since the gas is expected to cool all the way down to the cold phase, rather than be stuck at intermediate temperatures.

\subsection{Effects of heating model, MHD and resolution}
\begin{figure*}
	\centering
	\includegraphics[width=2.\columnwidth]{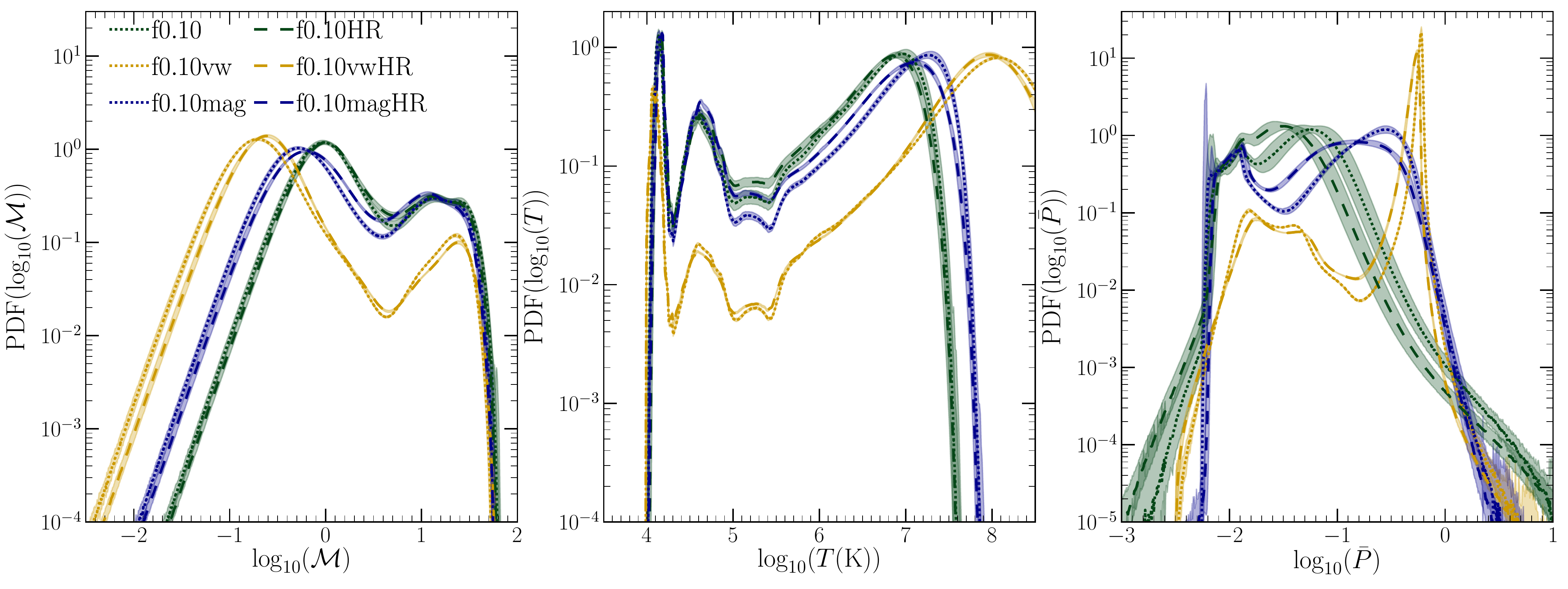}	
	
	\caption[mach_temp_pres_pdfs]{The volume-weighted PDFs of the logarithms of $\mathcal{M}$ (first column), temperature (second column) and pressure (third column) for runs with different heating models, MHD runs and higher resolution runs (dashed lines). Both magnetic fields and `vw' thermal heating lead to weaker turbulence. All the PDFs are averaged over 28 snapshots from $t=0.651$~$\mathrm{Gyr}$ to $t=1.003$~$\mathrm{Gyr}$.
	\label{fig:mach-temp-pres-PDF}
	}
\end{figure*}

\begin{figure*}
	\centering
	\includegraphics[width=2.\columnwidth]{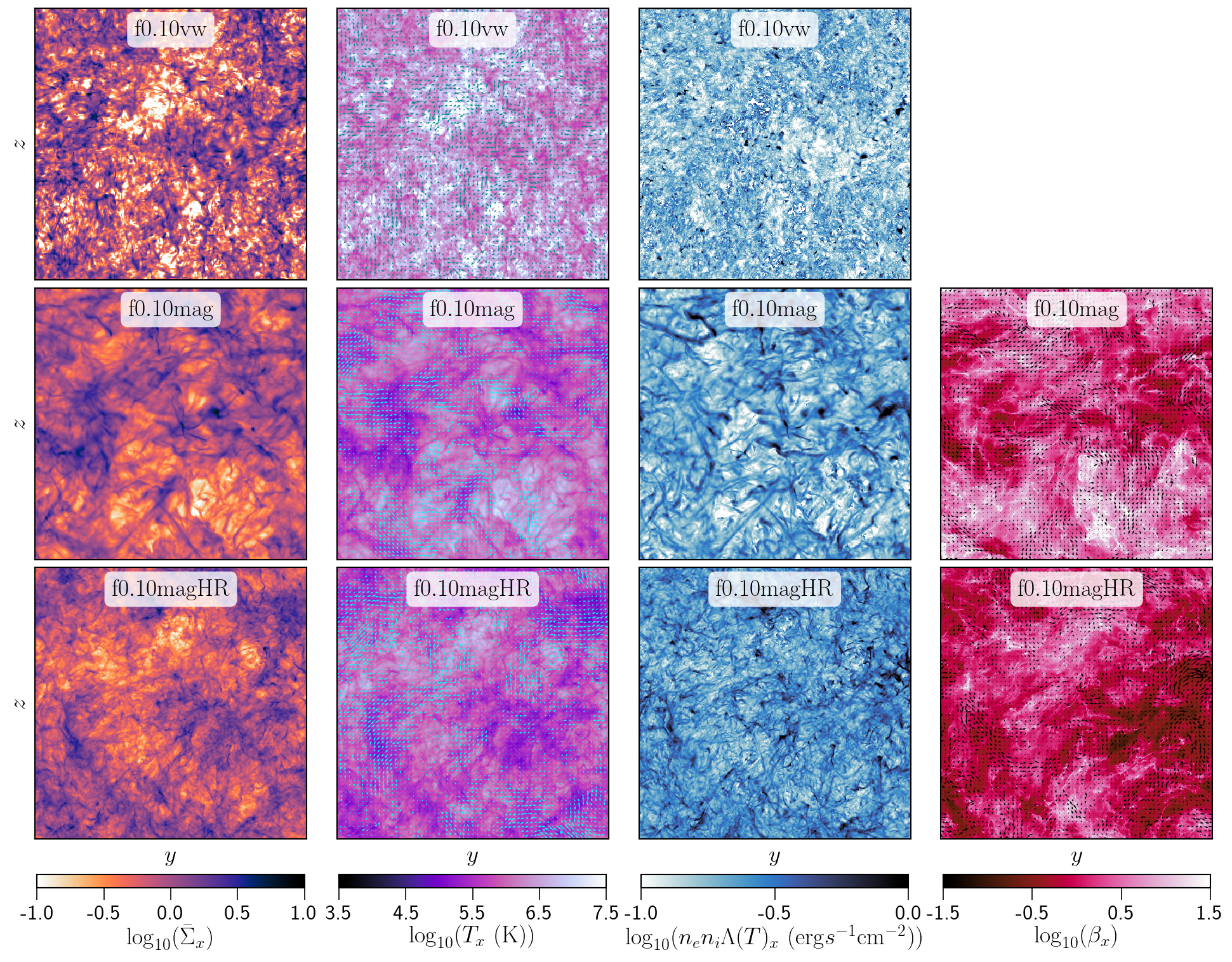}	
	
	\caption[projection_plots_MHD]{First three columns: same as \cref{fig:dens-temp-proj-2d}, but for runs f0.10vw, f0.10mag and f0.10magHR at $t=0.911$~$\mathrm{Gyr}$. The colorbar scale for column 3 has been adjusted. Fourth column: The volume-weighted projection of plasma beta, the ratio of thermal to magnetic pressure in log scale. The black arrows represent the projected $\mathbf{B}$ perpendicular to the LOS. The f0.10vw run has more small scale cold gas due to weaker turbulence. For the runs with magnetic fields, regions of low plasma beta are associated with gas in the cold phase. With increasing resolution, the size of the cold gas clumps becomes smaller.
	\label{fig:dens-temp-beta-proj-plot}
	}
\end{figure*}

In this subsection, we fix $f_{\mathrm{turb}}$ to the fiducial value ($0.10$) and discuss the effects of a different heating model ($Q_{\mathrm{vw}}$) and MHD. We then look at the effects of increasing resolution on these three runs.

\subsubsection{Effects of heating model}\label{subsubsec:diff_Q_model}
Here we present the effects of the heating model implemented (`mw' vs `vw' see \cref{subsubsec:glob_energy_balance}). We compare between the fiducial f0.10 run and the f0.10vw run. The mw heating model is an idealised version of density-dependent heating mechanisms such as heating by photons and cosmic rays. The vw heating model represents energy injection into the hot phase and its subsequent mixing with the rest of the gas through turbulence. This mimics the turbulent mixing of hot AGN-inflated bubbles with the ICM.

In \cref{fig:mach-temp-pres-PDF}, we show the $\mathcal{M}$ (first column), temperature (second column) and pressure (third column) PDFs for these two runs. The f0.10vw run has lower Mach numbers and less gas in the intermediate phase. This happens because by construction, most of the feedback thermal heat is added to the volume-filling hot phase. This reduces the net-cooling rate, since the hotter hot phase gas cools even slower. Because the turbulent heating rate is also proportional to the net cooling rate, these simulations have weaker turbulence, inefficient mixing and less intermediate temperature gas. Due to weaker turbulent smearing, the cold phase clouds are also much smaller in size, as seen in the first row of \cref{fig:dens-temp-beta-proj-plot}. In contrast, for the fiducial run, most of the thermal heat is added to the cold phase ($m_{\mathrm{cold}}/m_{\mathrm{tot}}\sim0.8$). This pushes the cold gas towards the fast-cooling intermediate phase, leading to an increase in the net-cooling rate and ultimately larger turbulent velocities with better mixing between the phases. The morphological features of the cold-phase gas, as well as the distribution of gas across different temperatures can be obtained from observations and be used to constrain the relative contribution from the different heating sources (for example, 
cosmic rays vs.~turbulent mixing) 
towards keeping the ICM hot.

\subsubsection{Effect of magnetic fields}\label{subsubsec:mag_field_compare}
The ICM is 
weakly magnetized, with plasma beta $\sim 100$ ($\beta=8\pi P/\mathbf{B}\cdot\mathbf{B}$) \citep{Carilli2002ARA&A,Govoni2004IJMPD,Bonafede2010A&A,Anderson2021PASA}. From Alfvèn’s flux freezing theorem, the field lines are frozen into the plasma and have to move along with it. Hence the compressed cold phase regions formed due to thermal instability are supposed to be magnetically dominated ($\beta<1$). This also leads to their filamentary structure \citep{Conselice2001AJ,Fabian2008Nature}. 

Magnetic fields also provide another channel of energy conversion in which turbulent kinetic energy is converted into magnetic energy. For the f0.10mag run, this leads to weaker turbulent density fluctuations and a lower cooling rate, which further decreases $\mathcal{M}_{\mathrm{hot}}$, as seen in the first column of \cref{fig:mach-temp-pres-PDF} (refer to column~5 in \cref{tab:sim_params} for its value). Because of the lower cooling rate, the hot phase is slightly hotter (second column) and its thermal pressure is higher. The cold phase  has low thermal pressure (peaking close to $P_{\mathrm{cutoff}}$), but it is partly supported by magnetic pressure. This is seen in the second and third rows of \cref{fig:dens-temp-beta-proj-plot}, where the cold, dense regions and their immediate surroundings are associated with low $\beta$ (fourth column). The higher magnetic pressure in the hot phase surrounding these cold phase regions may also contribute to larger values of $\sigma_{\ln\bar{P}}$ for the MHD runs in \cref{fig:hot-sig-mach}, as compared to predictions from the scaling relation.

\subsubsection{Effect of resolution}\label{subsubsec:res_compare}
Since we do not resolve $\ell_{\mathrm{cool}}$ (refer to \cref{subsec:2D_proj} for its definition), it is important to check the convergence of our results with resolution. 
In \cref{fig:mach-temp-pres-PDF}, we observe that on doubling the resolution (from $384^3$ to $768^3$), there is a slight increase in the hot phase Mach number (first column, low Mach number peak). The temperature (second column) PDFs also show more gas at intermediate temperatures and a lower temperature of the hot phase. In the pressure PDFs (third column), we find more gas at intermediate pressures between the two peaks corresponding to the phases. The hot phase peak is also at a slightly lower value. Most of these changes can be explained by an increase in the cooling rate, which leads to an increase in the strength of turbulence due to the energy balance condition in our setup. This leads to more mixing between the hot and cold phases and further decreases the temperature of the hot phase and transfers more gas into the intermediate phase. 

\cite{Mandelker2021arXiv} also find the cold gas mass fraction to increase with increasing resolution in their study of the multi-phase intergalactic medium. They argue that in simulations that do not resolve $\ell_{\mathrm{cool}}$, the gas piles up near $T\sim10^5~\mathrm{K}$ (corresponding to the peak of the cooling curve) and cooling to lower temperatures becomes inefficient. This is in agreement with our results. \cite{Fielding2020ApJ} show that upon increasing resolution, the bimodal distribution in pressure disappears for a radiative shear layer. Their highest resolution PDFs converge with the hot and cold phases in rough pressure equilibrium (see their fig.~5). However, \citet{Dutta2021} have found analytic cooling flow solutions around cold clouds that are sustained by mild pressure gradients between the ambient hot gas and the cold cloud. We find that our hot- and cold-phase pressure peaks move closer to each other upon increasing resolution, but this could partly be due to the increased strength of turbulent mixing (due to the larger cooling rate) in the higher-resolution runs.

Comparing the projection plots in the second and third rows of \cref{fig:dens-temp-beta-proj-plot}, we observe that the cold clouds have smaller physical size  for the higher resolution run. This can be seen quite clearly in the cold gas emission plot in the third column. These clouds are expected to become even smaller as we further increase resolution and reach convergence once the grid resolves important length scales. In addition to this, numerical diffusion is expected to affect the statistical properties of turbulence for scales $\ell\lesssim 30\Delta x$ \citep{Kitsionas2009A&A,federrath2010,Federrath2011ApJ}\footnote{Note that the power injected by driving turbulence may also leak to scales smaller than $L/2$, since the turbulent forcing power ($\rho v^3/\ell$) is coupled to the small-scale density variations \citep{Grete2017PhPl} in our multi-phase setup. This effect can further reduce the dynamical range of turbulence, since driving would affect statistics on scales smaller than $L/2$.}. Hence the condition for numerical convergence becomes: $\Delta x\lesssim \mathrm{min}(\ell_{\mathrm{cool}},\ell_{\mathrm{Field}},\ell_{\mathrm{\nu}})/30$, where $\ell_{\mathrm{Field}}=\sqrt{D t_{\mathrm{cool}}}$ corresponds to the scale where the thermal conduction time is equal to $t_{\mathrm{cool}}$, $D$ being the explicit diffusion constant \citep{koyama2004field,sharma2010thermal}. However, both $\ell_{\mathrm{Field}}$ and $\ell_\nu$ (the viscous length scale) are suppressed in the hot ICM \citep{Roberg-Clark2018PhRvL,Zhuravleva2019NatAs}, giving us even more stringent resolution requirements to get convergence for physical parameters. The scale $\ell_{\mathrm{\nu}}$ is also suppressed in the cold phase \citep{Li2020ApJ}. 

\section{Caveats and future work}\label{sec:caveats-future}
In this section, we discuss some of the shortcomings of our study and future prospects of our work.

The ICM is known to be stratified due to the gravitational potential set by the dark matter halo. However, we have ignored the effects of stratification in this set of idealised simulations. Density fluctuations in stratified turbulence are supposed to be much stronger \citep{Mohapatra2020,Mohapatra2021MNRAS} and comparable to the magnitudes seen in \cref{fig:hot-sig-mach}. They can also influence the mode of density perturbations in the ICM, changing them from adiabatic to isobaric. 
The 
dense cold-phase gas in our simulations would 
fall in the presence of gravity, as seen in \cite{Wang2021MNRAS}. Our runs with low $f_{\mathrm{turb}}$ produce a somewhat hotter hot phase. But in the presence of stratification, this thermally heated gas can rise against the density gradient, expand and cool adiabatically. The ratio between the cooling time and the free-fall time is known to play an important role in thermal instability in these environments \citep{sharma2012thermal,choudhury2016,Choudhury2019,Voit2021ApJ}. 
We plan to include stratification and study its effects in a future thermal instability study.

In order to directly focus on the impact of different levels of turbulent heating, we fixed our initial conditions (density and temperature) across all simulations. While the different levels of turbulent to thermal pressure ratio can represent the gas in halos of different masses (lower mass halos being more turbulent), our initial conditions are more applicable to cool and thermally unstable regions in groups and clusters. 

We have conducted two MHD simulations, with fixed initial magnetic field geometry. We have studied their effects on the gas distribution across different temperatures, densities and pressure. However, our results could be sensitive to the choice of the initial field configuration and its initial amplitude, which are not well-constrained from observations. We plan to investigate these effects in a follow-up study focusing on the effects of the magnetic field geometry, amplitude and orientation, on the structure and distribution of the multi-phase gas.

In order to make accurate predictions of different observables at different temperatures, such as the column densities of H\texttt{I}, Mg\texttt{II}, C\texttt{IV}, O\texttt{VI}, Fe\texttt{XXVI}, etc., we need to track these species through chemical networks, non-equilibrium ionisation and photoionisation modelling. However, these methods are computationally expensive and have not been included in this study.

We have not included thermal conduction in our simulations. Thermal conduction is supposed to be suppressed and anisotropic in the ICM plasma \citep{roberg-clark2016,Meinecke2021arXiv}. But as shown in \cite{gaspari2013constraining}, it can wipe out small scale structures even when suppressed by a factor of $\sim10$ relative to the Spitzer value. It can also populate gas at intermediate temperatures and introduce isothermal modes of perturbation. 

Our local ICM boxes only have a resolution of $\approx100$--$50$~$\mathrm{pc}$ for our standard and high resolution runs. We require much higher spatial resolution to resolve mixing layers in the multiphase gas 
\citep{Fielding2020ApJ}. The scale separation between the two phases may also be much more than what we observe in this study -- we observe that the size of cold clouds decreases on increasing resolution.

\section{Conclusions}\label{sec:Conclusion}
In this study we have focused on the effects of different fractions of turbulent heating in an energetically balanced simulation of the ICM. Here we summarise some of the main takeaway points of our work:
\begin{enumerate}
    \item Turbulence seeds thermal instability in the hot phase gas, which separates into hot and cold phases. These phases are distinguishable as separate peaks in the Mach number, temperature, pressure and density PDFs. Most of the mass of the gas is in the cold phase but the hot phase occupies most of the volume.
    \item Mixing due to turbulence increases the fraction of gas at the thermally unstable intermediate regions near the peak of the cooling curve ($T\sim10^{5.5}$~$\mathrm{K}$). 
    \item The density fluctuations in the thermally unstable hot phase gas are much larger than the fluctuations predicted by $\sigma_s$--$\mathcal{M}$ scaling relations based on homogeneous idealised turbulence. But the pressure fluctuations in the hot phase are unaffected by thermal instability and they obey the same scaling relations. 
    \item The mode of density fluctuations in the hot phase changes from isobaric to adiabatic for stronger turbulence. The phase transition from the hot phase to the cold phase during 
    condensation is mostly isochoric. The fluctuations in the cold phase are isothermal at the cooling-cutoff temperature.
    \item The intermediate temperature phase gas is the most unstable gas phase, but its stability increases with increasing turbulent velocities.
    \item Using different heating prescriptions (volume-weighted vs mass-weighted) can affect the amount of gas in the two phases. The volume-weighted prescription deposits more feedback heat in the hot volume-filling phase.
    \item In MHD runs, turbulent kinetic energy is converted into magnetic energy, which leads to lower turbulent velocities. The gas in the cold phase is at low thermal pressure and is dominated by magnetic pressure. 
\end{enumerate}

\section*{Acknowledgements}
This work was carried out during the ongoing COVID-19 pandemic. The authors would like to acknowledge the health workers all over the world for their role in fighting in the frontline of this crisis. The authors would like to thank Dr.~Philipp Grete for a constructive referee report, which helped to improve this work. RM thanks Prof.~Naomi M.~McClure-Griffiths for organising a writing retreat, where a part of this work was written. CF acknowledges funding provided by the Australian Research Council (Future Fellowship FT180100495), and the Australia-Germany Joint Research Cooperation Scheme (UA-DAAD). PS acknowledges a Swarnajayanti Fellowship (DST/SJF/PSA-03/2016-17) and a National Supercomputing Mission (NSM) grant from the Department of Science and Technology, India. We further acknowledge high-performance computing resources provided by the Leibniz Rechenzentrum and the Gauss Centre for Supercomputing (grants~pr32lo, pr48pi and GCS Large-scale project~10391), the Australian National Computational Infrastructure (grant~ek9) in the framework of the National Computational Merit Allocation Scheme and the ANU Merit Allocation Scheme. The simulation software FLASH was in part developed by the DOE-supported Flash Center for Computational Science at the University of Chicago.

\section{Data Availability}
All the relevant data associated with this article is available upon request to the corresponding author.

\section{Additional Links}
Movies of projected density and temperature of different simulations are available as online supplementary material, as well as at the following links on youtube:
\begin{enumerate}
    \item \href{https://youtu.be/F8QsytVrsws}{Movie} of the fiducial simulation.
    \item \href{https://youtu.be/-_0WIfRL9tI}{Movie} of the f0.10HR simulation.
    \item \href{https://youtu.be/FJnlmVOgQMQ}{Movie} of the f0.10vwHR simulation.
    \item \href{https://youtu.be/NUvp18fKETM}{Movie} of the f0.10magHR simulation.
\end{enumerate}
We also show a \href{https://youtu.be/OtsIaaD6I5Q}{movie} of \cref{fig:pres-dens-mp-2DPDF} and a \hyperlink{https://youtu.be/3tcXW9OH6F8}{movie} of tracer particles.
\section{Software used}
We have used the following software and packages for our work:
FLASH \citep{Fryxell2000,Dubey2008}, matplotlib \citep{Hunter4160265}, cmasher \citep{Ellert2020JOSS}, scipy \citep{Virtanen2020}, NumPy \citep{Harris2020}, h5py \citep{collette_python_hdf5_2014} and astropy \citep{astropy2018}.


\bibliographystyle{mnras}
\bibliography{refs.bib} 



\bsp	
\label{lastpage}
\end{document}